\documentclass[
aps,
pra,
twocolumn,
superscriptaddress,
floatfix,
longbibliography
]{revtex4-1}
\usepackage[dvipdfmx]{graphicx}
\usepackage{dcolumn}
\usepackage{bm}
\usepackage{ulem}
\usepackage{amsmath}
\usepackage{amssymb}
\usepackage{txfonts}
\usepackage{hyperref}
\usepackage{color} 
\usepackage{xcolor}
\usepackage{listings}
\usepackage{cancel}
\newcommand{\bs}   {\boldsymbol}

\newcommand{\e}{{\rm e}}
\newcommand{\imag}{{\rm i}}

\hypersetup{
  colorlinks=true,
  linkcolor=[rgb]{0.60,0.00,0.00},
  citecolor=[rgb]{0.00,0.00,0.60},
  urlcolor=[rgb]{0.00,0.00,0.60},
  setpagesize=false
}

\definecolor{codegreen}{rgb}{0,0.6,0}
\definecolor{codegray}{rgb}{0.5,0.5,0.5}
\definecolor{codepurple}{rgb}{0.58,0,0.82}
\definecolor{backcolour}{rgb}{0.95,0.95,0.92}

\lstdefinestyle{mystyle}{
    backgroundcolor=\color{backcolour},   
    commentstyle=\color{codegreen},
    keywordstyle=\color{magenta},
    numberstyle=\tiny\color{codegray},
    stringstyle=\color{codepurple},
    basicstyle=\ttfamily\footnotesize,
    breakatwhitespace=false,         
    breaklines=true,                 
    captionpos=b,                    
    keepspaces=true,                 
    numbers=left,                    
    numbersep=5pt,                  
    showspaces=false,                
    showstringspaces=false,
    showtabs=false,                  
    tabsize=2
}

\begin{document}

\title{
  Gutzwiller wave function on a quantum computer
  using a discrete Hubbard-Stratonovich transformation 
}

\author{Kazuhiro~Seki}
\affiliation{Quantum Computational Science Research Team, RIKEN Center for Quantum Computing (RQC), Saitama 351-0198, Japan}

\author{Yuichi~Otsuka}
\affiliation{Quantum Computational Science Research Team, RIKEN Center for Quantum Computing (RQC), Saitama 351-0198, Japan}
\affiliation{Computational Materials Science Research Team, RIKEN Center for Computational Science (R-CCS),  Hyogo 650-0047,  Japan}

\author{Seiji~Yunoki}
\affiliation{Quantum Computational Science Research Team, RIKEN Center for Quantum Computing (RQC), Saitama 351-0198, Japan}
\affiliation{Computational Materials Science Research Team, RIKEN Center for Computational Science (R-CCS),  Hyogo 650-0047,  Japan}
\affiliation{Computational Quantum Matter Research Team, RIKEN Center for Emergent Matter Science (CEMS), Saitama 351-0198, Japan}
\affiliation{Computational Condensed Matter Physics Laboratory, RIKEN Cluster for Pioneering Research (CPR), Saitama 351-0198, Japan}

\begin{abstract}
  We propose a quantum-classical hybrid scheme 
  for implementing the nonunitary Gutzwiller factor
  using a discrete Hubbard-Stratonovich transformation, 
  which allows us to express the Gutzwiller factor as a
  linear combination of unitary operators involving
  only single-qubit rotations, at the cost of
  the sum over the auxiliary fields. 
  To perform the sum over the auxiliary fields,
  we introduce two approaches that have complementary features.
  The first approach employs a linear-combination-of-unitaries circuit, 
  which enables one to probabilistically prepare the Gutzwiller wave function on a quantum computer, while
  the second approach uses importance sampling to estimate observables stochastically, similar
  to a quantum Monte Carlo method in classical computation. 
  The proposed scheme is demonstrated with numerical simulations
  for the half-filled Fermi-Hubbard model.
  Furthermore, we perform quantum simulations using a real quantum device,
  demonstrating that the proposed scheme can reproduce
  the exact ground-state energy of the two-site Fermi-Hubbard model within error bars.
\end{abstract}

\date{\today}

\maketitle

\section{Introduction}

Solving quantum many-body systems directly using classical computers requires
exponentially large computational resources 
that are often beyond the feasibility of current high-performance
computing facilities. To overcome this difficulty, at least in part, a tremendous amount of effort has been 
devoted so far and several 
theoretical and numerical techniques have been successfully developed~\cite{Fehske_book}. 
It should also be pointed out that, as one of the next-generation 
computing paradigms,
quantum computing~\cite{Feynman1982} 
has attracted growing interest for solving quantum many-body systems, 
which is becoming realistic, evidenced by the recent technological advances~\cite{Nakamura1999,optical_RMP2007,Ladd2010,RevModPhys.85.623,Chow2014,Barends2014,Riste2015,Kelly2015,Arute2019,Zhong2020}. 
In this regards,
the variational-quantum-eigensolver 
method and its variants have been proposed and demonstrated
for computing
ground-state~\cite{Yung2014,Peruzzo2014,Wecker2015vqe,O'Malley2016,McClean2016,Kandala2017,Li2017,Mazzola2019,arute2020hartreefock,Suchsland2021,Stanisic2021} and
low-lying excited-state~\cite{McClean2017,Colless2018,Parrish2019,nakanishi2018subspacesearch,heya2019subspace,huggins2019nonorthogonal} properties
of quantum-many-body systems by exploiting noisy intermediate-scale quantum (NISQ)~\cite{Preskill2018} computers
and classical computers in a hybrid manner. 
For recent reviews on variational quantum algorithms,
see for example Refs.~\cite{McArdle2020,Endo2021,Cerezo2021,tilly2021variational}. 
It is also remarkable that
physically motivated wave function
such as Gutzwiller- and Jastrow-type wave functions~\cite{Mazzola2019}
and a resonating-valence-bond-type wave function~\cite{seki2020vqe} have
been implemented with NISQ computers.

The Gutzwiller wave function 
is known as a variational state for
quantum-many-body systems in condensed matter physics 
that allows us to take into account electron correlation effects
beyond the level of a single Slater-determinant state~\cite{Gutzwiller1963}.
Despite its formal simplicity,
the Gutzwiller-type wave functions, 
including
Gutzwiller-projected Fermi-sea states~\cite{Vollhard1984}, 
Gutzwiller-projected BCS states~\cite{Anderson1987,Himeda2000}, and
Gutzwiller-projected Hartree-Fock states~\cite{Ogata2003}, 
can describe ground and low-lying excited states of several quantum many-body systems
such as a lattice model of dimers~\cite{Fabrizio2007}, 
the Haldane-Shastry model~\cite{Haldane1988,Shastry1988}, and 
$t$-$J$-type models~\cite{Yokoyama1987tJ,Kuramoto1991,Yokoyama1991,Himeda2002,Yunoki2005,Lee2006,Yunoki2006BCS} 
(for a recent ground-state phase diagram of the $t$-$t'$-$J$ model
using the density-matrix-renormalization-group method, see Ref.~\cite{Jiang2021})
qualitatively or even exactly in some particular cases. 

In classical computation, 
the Gutzwiller wave function can be implemented
rather straightforwardly when 
the Gutzwiller factor is diagonal in a computational basis.
In quantum computation, 
one can also choose the computational basis states
so that the Gutzwiller factor is diagonal. 
However, 
implementing a Gutzwiller-type wave function on a quantum computer 
is not straightforward due to its nonunitarity,  
and several schemes for implementing it
have been developed~\cite{Mazzola2019,Yao2021,Murta2021}.
Mazzola {\it et al.}~\cite{Mazzola2019} evaluates
the expectation value of energy with respect to
a Jastrow-type wave function~\cite{Jastrow1955,Capello2005} by 
measuring
the transformed Hamiltonian
$\hat{P}_{\rm J} \hat{\cal H} \hat{P}_{\rm J}$
and the squared Jastrow factor 
$(\hat{P}_{\rm J})^2$ 
with a suitably truncated expansion of
the Jastrow factor $\hat{P}_{\rm J}$. 
Murta and Fern\'{a}ndez-Rossier~\cite{Murta2021} proposed
a quantum circuit for 
probabilistically preparing the Gutzwiller wave function
using ancillary qubits.  
It is also noteworthy that, aside from the Gutzwiller wave function,
general frameworks for probabilistically performing nonunitary operations
on a quantum computer have been proposed~\cite{Gingrich2004,Liu2021,kosugi2021probabilistic}.

In this paper, we propose another scheme for implementing
the Gutzwiller wave function using
a discrete version~\cite{Hirsch1983} of
the Hubbard-Stratonovich transformation~\cite{Hubbard1959}, 
which allows us to represent the Gutzwiller factor
as a linear combination of unitary operators,
at the expense of introducing the auxiliary fields. 
In order to sum all the auxiliary fields,
we introduce two different but complimentary approaches based on  
(i) a quantum circuit for the linear combination of unitary operators and
(ii) an importance sampling technique. 
Furthermore, the proposed scheme is demonstrated using numerical simulations as well as 
a real quantum device.

The rest of this paper is organized as follows.
In Sec.~\ref{sec:form},
we provide formalism of the proposed scheme. 
We first define the Hamiltonian of the Fermi-Hubbard model and
the Gutzwiller wave function.
Then we describe the discrete Hubbard-Stratonovich
transformation for the Gutzwiller factor, and
introduce the Jordan-Wigner transformation 
to construct concrete quantum circuits 
for implementing the Gutzwiller wave function
on a quantum computer. 
In Sec.~\ref{sec:sum},
we describe two complementary approaches 
for performing the sum over the auxiliary fields.
The first approach employs a linear-combination-of-unitaries circuit
for probabilistically preparing the Gutzwiller wave function on a quantum computer. 
The second approaches uses an importance sampling technique 
to stochastically evaluate observables with respect to the
Gutzwiller wave function.
We also describe a simplification scheme that is applicable
when the trial state is a separable state
with respect to the spin degrees of freedom.
The two approaches are demonstrated by 
numerical simulations for the Fermi-Hubbard model up to $12$ sites.
In Sec.~\ref{sec:results},
we apply the proposed scheme for
calculating ground-state properties of the 
two-site Fermi-Hubbard model at half filling
by using a NISQ computer. 
First, we summarize the Gutzwiller wave function approach
for the ground-state of two-site Fermi-Hubbard model at half filling,
where the Gutzwiller wave function can describe the
ground state exactly. 
Then we show the results obtained by 
using a NISQ computer. 
Conclusions and discussions are given in Sec.~\ref{sec:conclusions}. 
In Appendix~\ref{app:A},
we provide a general scheme for finding discrete
Hubbard-Stratonovich transformations, 
which decompose an exponentiated density-density 
interaction term into a linear combination of two-qubit unitary
operators.
In Appendix~\ref{sec:pp},
we prove the absence of the phase problem
in the second approach for the Fermi-Hubbard model
on a bipartite lattice at half filling.

\section{Model and Formalism}\label{sec:form}

\subsection{Fermi-Hubbard model}\label{sec:model}

We consider the Fermi-Hubbard model
defined by the Hamiltonian 
\begin{equation}
  \hat{\cal H}
  =\hat{K} + U\hat{D},  
  \label{eq:ham}
\end{equation}
where 
\begin{equation}
  \hat{K} 
  = -J \sum_{\sigma=\uparrow,\downarrow}\sum_{\langle i,j\rangle}
  \left(
  \hat{c}_{i\sigma}^\dag
  \hat{c}_{j\sigma}
  +{\rm H.c.}
  \right)
  \label{eq:K}
\end{equation}
and
\begin{equation}
  \hat{D}
  = \sum_{i=1}^{N_{\rm site}}
    \left(\hat{n}_{i\uparrow}-\tfrac{1}{2}\right)
    \left(\hat{n}_{i\downarrow}-\tfrac{1}{2}\right). 
    \label{eq:D}
\end{equation}
Here, 
$J$ is the hopping parameter,
$U$ is the interaction parameter, 
$N_{\rm site}$ is the number of sites, and 
$\hat{c}_{i\sigma}^\dag$ $(\hat{c}_{i\sigma})$
is the creation (annihilation) operator of a fermion at 
site $i\, (=1,2,\cdots,N_{\rm site})$ with spin $\sigma\, (=\uparrow,\downarrow)$. 
The summation $\sum_{\langle i,j \rangle}\cdots$ denotes the sum over all pairs of nearest-neighbor sites $i$ and $j$.  
$\hat{n}_{i\sigma}=\hat{c}_{i\sigma}^\dag \hat{c}_{i\sigma}$
is the fermion number operator at site $i$ with spin $\sigma$.
In this study, we assume $J>0$ and $U\geqslant 0$.

\subsection{Gutzwiller wave function}

The Gutzwiller wave function 
\begin{equation}
  |\psi_g\rangle \equiv
\frac{
  \e^{-g\hat{D}}|\psi_0\rangle}
{\sqrt{\langle\psi_0 |
    \e^{-2g\hat{D}}
    |\psi_0 \rangle}},
\label{eq:Gutzwiller}
\end{equation}
is known as a variational state for the Fermi-Hubbard model~\cite{Gutzwiller1963}. 
Here, $ \e^{-g\hat{D}}$ is the Gutzwiller factor with  
$0 \leqslant g < \infty$ being the dimensionless variational parameter that 
penalizes the double occupancy of fermions at the same site, and
$|\psi_0\rangle$ is a trial state. 
In this study, we assume that $|\psi_0\rangle$
is normalized as $\langle \psi_0|\psi_0\rangle=1$ and 
it is an eigenstate of the total particle-number operator
$\hat{N}\equiv\sum_{i}\sum_{\sigma}\hat{n}_{i\sigma}$.
Since $[\hat{N},\hat{D}]\equiv\hat{N}\hat{D}-\hat{D}\hat{N}=0$, 
$|\psi_g\rangle$ is also an eigenstate of $\hat{N}$. 
Typically, 
the trial state $|\psi_0\rangle$ is chosen as
a single Slater-determinant state
such as the ground state of $\hat{K}$ or
a single-particle mean-field Hamiltonian. 
The Gutzwiller wave function $|\psi_g\rangle$ can take into account
electron correlation effects beyond the trial state $|\psi_0 \rangle$.

There is one remark on the Gutzwiller factor. 
Originally, the Gutzwiller factor was
introduced in the following form~\cite{Gutzwiller1963} 
\begin{equation}
  \hat{G}(\tilde{g})\equiv\prod_{i=1}^{N_{\rm site}}\left[1-(1-\tilde{g})\hat{n}_{i\uparrow}\hat{n}_{i\downarrow}\right]
  \overset{\tilde{g}\not=0}
  =\tilde{g}^{\sum_i \hat{n}_{i\uparrow}\hat{n}_{i\downarrow}}, 
  \label{eq:G}
\end{equation}
where $0\leqslant \tilde{g} \leqslant 1$ is the variational parameter
and the right-hand side is valid for $\tilde{g}\not =0$.
If $\tilde{g}=0$,
$\hat{G}(0)=\prod_{i=1}^{N_{\rm site}}(1-\hat{n}_{i\uparrow}\hat{n}_{i\downarrow})$, which excludes fermion configurations
with doubly occupied sites from $|\psi_0\rangle$, and
$\hat{G}(0)$ is called the Gutzwiller projector~\cite{Gebhard_book}.
Provided that $|\psi_0\rangle$ is an eigenstate of the total particle-number operator $\hat{N}$,  
we can easily show that the following equality holds: 
\begin{equation}
|\psi_g\rangle =
\frac{\e^{-g\hat{D}}|\psi_0\rangle}
     {\sqrt{\langle\psi_0 |
    \e^{-2g\hat{D}}
    |\psi_0 \rangle}}
     =
\frac{
  \hat{G}(\tilde{g})|\psi_0\rangle}
     {\sqrt{\langle\psi_0 |
         \hat{G}(\tilde{g})^2
    |\psi_0 \rangle}} 
\end{equation}
with the parameters $\tilde{g}$ and $g$ satisfying the relation (see, e.g., Ref.~\cite{Yokoyama1987})
\begin{equation}
  \tilde{g}=\e^{-g}.  
  \label{eq:tilde_g}
\end{equation}
Therefore, 
using $\e^{-g\hat{D}}$ 
is equivalent to
using $\hat{G}(\tilde{g})$ for expressing the Gutzwiller wave function $|\psi_g\rangle$, 
despite that $\e^{-g\hat{D}}\not=\hat{G}(\tilde{g})$.  
The reason why we use $\e^{-g\hat{D}}$ is 
simply because $\e^{-g\hat{D}}$ is readily compatible with
the Hubbard-Stratonovich transformation, 
as described in the next section.

\subsection{Discrete Hubbard-Stratonovich transformation}~\label{sec:HST}

To express the Gutzwiller factor $\e^{-g\hat{D}}$ as
a linear combination of unitary operators,
we introduce the discrete Hubbard-Stratonovich
transformation~\cite{Hirsch1983}
\begin{equation}
  \e^{-g
      \left(\hat{n}_{i\uparrow}-\tfrac{1}{2}\right)
      \left(\hat{n}_{i\downarrow}-\tfrac{1}{2}\right)
  }
  = \gamma \sum_{s_i=\pm 1} \e^{\imag \alpha s_i
      \left(\hat{n}_{i\uparrow}+\hat{n}_{i\downarrow}-1 \right)
  },
  \label{eq:HST}
\end{equation}
where
 $s_{i}\,(=\pm 1)$ is the discrete auxiliary field, 
$\gamma=\e^{g/4}/2$, and
\begin{equation}
  \alpha=\arccos{\left(\e^{-g/2}\right)}.
  \label{eq:alpha}
\end{equation}
Since the fermion number operators commute with each other,
$[\hat{n}_{i\sigma},\hat{n}_{j\sigma^\prime}]=0$,
the Gutzwiller factor can be written simply as
\begin{alignat}{1}
  \e^{-g\hat{D}} &=
  \gamma^{N_{\rm site}}
  \prod_{i=1}^{N_{\rm site}} 
  \sum_{s_i=\pm 1}
  \e^{2\imag \alpha s_i \hat{\eta}_{i}^z }
  \label{eq:gf} \\
  &=
  \gamma^{N_{\rm site}}
  \prod_{i=1}^{N_{\rm site}} 
  \left(
  \e^{2\imag \alpha \hat{\eta}_{i}^z }
  +
  \e^{-2\imag \alpha \hat{\eta}_{i}^z }
  \right),  
  \label{eq:HST2}
\end{alignat}
where 
\begin{equation}
  \hat{\eta}_{i}^z \equiv \frac{1}{2}\left(\hat{n}_{i\uparrow} +\hat{n}_{i\downarrow}-1 \right)
  \label{eq:eta}
\end{equation}
is introduced to simplify the notation. 
Now the nonunitary Gutzwiller factor 
[left-hand side of Eq.~(\ref{eq:HST2})]
is expressed as a linear combination of unitary operators 
[right-hand side of Eq.~(\ref{eq:HST2})] 
after summing all the auxiliary fields $\{s_i\}_{i=1}^{N_{\rm site}}$. 
The different decomposition schemes of the Hubbard-Stratonovich
transformation that are potentially useful for other purposes are provided in Appendix~\ref{app:A}.

The expectation value of an operator $\hat{O}$ with respect to
the Gutzwiller wave function $|\psi_g \rangle$ is given by
\begin{alignat}{1}
  \langle \hat{O}\rangle
  &\equiv \langle \psi_g |\hat{O} |\psi_g \rangle
  =
  \frac
  {\langle\psi_0 |
    \e^{-g\hat{D}}
    \hat{O}
    \e^{-g\hat{D}}
    |\psi_0 \rangle }
  {\langle\psi_0 |
    \e^{-2g\hat{D}}
    |\psi_0 \rangle }
  \label{eq:aveO} \\
    &=\frac
          {
            \sum_{\bs{s}}
            \langle\psi_0 |
            \prod_{i=1}^{N_{\rm site}}            
            \e^{2 \imag \alpha s_{i,2} \hat{\eta}_i^z}
            \hat{O}
            \prod_{j=1}^{N_{\rm site}}            
            \e^{2 \imag \alpha s_{j,1} \hat{\eta}_j^z}
            |\psi_0 \rangle
          }
  {
    \sum_{\bs{s}^\prime}
    \langle\psi_0 |
    \prod_{i=1}^{N_{\rm site}}            
    \e^{2 \imag \alpha (s_{i,1}^\prime+s_{i,2}^\prime) \hat{\eta}_i^z}
    |\psi_0 \rangle}, 
  \label{eq:O}
\end{alignat}
where
\begin{equation}
  \sum_{\bs{s}} \cdots \equiv \prod_{\tau=1}^{N_{\tau}} \prod_{i=1}^{N_{\rm sites}}
  \sum_{s_{i,\tau}=\pm 1}\cdots
  \label{eq:sum}
\end{equation}
represents the sum over the auxiliary fields $\{ \{s_{i,\tau}=\pm1\}_{i=1}^{N_{\rm site}} \}_{\tau=1}^{N_\tau}$ with $N_\tau=2$, 
implying that the total number of terms in the sum over $\bs{s}$ is $2^{N_\tau \cdot N_{\rm site}}=4^{N_{\rm site}}$.
Notice that another label $\tau\, (=1,2)$ for the auxiliary fields
is introduced to distinguish the auxiliary fields
corresponding to the bra and ket states, i.e., 
$\tau=1$ for $\langle \psi_g|$ and $\tau=2$ for $|\psi_g\rangle$~\cite{note_decomp}. 

\subsection{Jordan-Wigner transformation}

The formalism described above is given in terms of the fermion operators.
In order to implement the proposed scheme on a quantum computer, 
we now apply the Jordan-Wigner transformation of the form 
\begin{equation}
  \hat{c}_{i\sigma}^\dag
  \overset{\rm JWT}{=}
  \frac{1}{2}(\hat{X}_{i_\sigma} - \imag \hat{Y}_{i_\sigma})
  \prod_{k < {i_\sigma}} \hat{Z}_k
  \label{eq:JWT1} 
\end{equation}
and
\begin{equation}
  \hat{c}_{i\sigma}
  \overset{\rm JWT}{=}
  \frac{1}{2}(\hat{X}_{i_\sigma} + \imag \hat{Y}_{i_\sigma})
  \prod_{k < {i_\sigma}} \hat{Z}_k, 
  \label{eq:JWT2}
\end{equation}
where $i_{\sigma}\, (=1,2,\cdots,2N_{\rm site})$
is the one-dimensional label for
the site and spin indexes 
and
$\hat{X}_{i_\sigma}$, 
$\hat{Y}_{i_\sigma}$, and
$\hat{Z}_{i_\sigma}$ are
Pauli $X$, $Y$, and $Z$ operators acting on the $i_\sigma$th qubit. 
By $\cdots \overset{\rm JWT}{=} \cdots$ in Eqs.~(\ref{eq:JWT1}) and (\ref{eq:JWT2}), 
we denote that the fermion operators on the left-hand side
are expressed in terms of the Pauli operators under
the Jordan-Wigner transformation.

Using the Jordan-Wigner transformation,  
the kinetic term in Eq.~(\ref{eq:K}) 
and the interaction term in Eq.~(\ref{eq:D}) 
of the Hamiltonian can be expressed with the Pauli operators as 
\begin{equation}
  \hat{K}
  \overset{\rm JWT}{=}
  -\frac{J}{2} \sum_\sigma \sum_{\langle i_\sigma, j_\sigma \rangle}
  \left(
  \hat{X}_{i_\sigma} \hat{X}_{j_\sigma} +
  \hat{Y}_{i_\sigma} \hat{Y}_{j_\sigma}
  \right)
  \hat{Z}_{{\rm JW},i_\sigma j_\sigma}
  \label{eq:JWK}
\end{equation}
and
\begin{equation}
  \hat{D}
  \overset{\rm JWT}{=}
  \frac{1}{4} \sum_{i=1}^{N_{\rm site}} \hat{Z}_{i_\uparrow} \hat{Z}_{i_\downarrow},
  \label{eq:JWD}
\end{equation}
respectively, where
$\hat{Z}_{{\rm JW},ij}=\prod_{i \lessgtr k \lessgtr j} \hat{Z}_k$
is the Jordan-Wigner string for $i \lessgtr k \lessgtr j$
and $\hat{Z}_{{\rm JW},ij}=\hat{I}$ (identity operator) for $i=j\pm 1$.
Similarly, the operator $\hat{\eta}_{i}^z$ in Eq.~(\ref{eq:eta}) can be expressed as 
\begin{equation}
   \hat{\eta}_{i}^z
  \overset{\rm JWT}{=}
  -\frac{1}{4}\left(\hat{Z}_{i_\uparrow}+\hat{Z}_{i_\downarrow}\right)
  \overset{\cdot}{=}
  \frac{1}{2}
  \begin{bmatrix}
    -1 & 0 & 0 & 0 \\
    0 & 0 & 0 & 0 \\
    0 & 0 & 0 & 0 \\
    0 & 0 & 0 & 1
  \end{bmatrix},
\end{equation}
where $\overset{\cdot}{=}$ indicates the matrix representation and 
the matrix here is represented with the computational basis states
$|00\rangle \equiv |0\rangle_{i_\uparrow} |0\rangle_{i_\downarrow}$,
$|01\rangle \equiv |0\rangle_{i_\uparrow} |1\rangle_{i_\downarrow}$,
$|10\rangle \equiv |1\rangle_{i_\uparrow} |0\rangle_{i_\downarrow}$, and 
$|11\rangle \equiv |1\rangle_{i_\uparrow} |1\rangle_{i_\downarrow}$ 
with $\hat{Z}_{i_{\sigma}}|0\rangle_{i_\sigma}=|0\rangle_{i_\sigma}$ and 
$\hat{Z}_{i_{\sigma}}|1\rangle_{i_\sigma}=-|1\rangle_{i_\sigma}$. 
Finally, the rotation generated by
$2\hat{\eta}_{i}^z$ can be expressed
as a product of the one-qubit rotations, i.e., 
\begin{equation}
  \e^{2\imag \alpha \hat{\eta}_{i}^z}
  \overset{\rm JWT}{=}
  \hat{R}_{Z_{i_\uparrow}}\left(\alpha \right)
  \otimes
  \hat{R}_{Z_{i_\downarrow}}\left(\alpha \right)
  \overset{\cdot}{=}
  \begin{bmatrix}
    \e^{-\imag \alpha} & 0 & 0 & 0 \\
    0 & 1 & 0 & 0 \\
    0 & 0 & 1 & 0 \\
    0 & 0 & 0 & \e^{\imag \alpha}
  \end{bmatrix},
  \label{eq:etarot}
\end{equation}
where
\begin{equation}
  \hat{R}_Z(\alpha)=\exp{\left(-\imag \alpha \hat{Z}/2\right)}.
\end{equation}
Notice that  
$\e^{2\imag \alpha \hat{\eta}_{i}^z}$
acts nontrivially only on the empty state (corresponding to $|00\rangle$) and
the doubly occupied state (corresponding to $|11\rangle$). 
Equation~(\ref{eq:etarot}) shows that, under the
Jordan-Wigner transformation,
$\e^{2\imag \alpha \hat{\eta}_{i}^z}$ is
expressed simply as a direct product of
the single-qubit $Z$ rotation gates with the same
rotation angle $\alpha$.

\section{Sum over auxiliary fields}\label{sec:sum}

Since the terms involved in the sum over the auxiliary fields in Eqs.~(\ref{eq:gf}) and (\ref{eq:O}) 
increases exponentially in $N_{\rm site}$,
performing directly 
the sum becomes unfeasible
as $N_{\rm site}$ is large.
Nevertheless, here we introduce two approaches, based on 
(i) a quantum circuit for a linear combination of unitary operators and
(ii) an importance sampling technique, 
for performing the sum over the auxiliary fields. 
These two approaches have complementary features.  

The first approach based on a quantum circuit for a linear combination of unitary operators 
allows us to probabilistically prepare
the Gutzwiller wave function $|\psi_g\rangle$ on a quantum computer
by using $N_{\rm site}$ ancillary qubits, which can be trivially reduced to one if an ancillary qubit is reused, 
and $2N_{\rm site}$ controlled-$R_Z$ operations. 
However, the probability for successfully preparing the
desired state decreases exponentially in $N_{\rm site}$. 
In the second approach based on an importance sampling technique, 
the expectation values of observables are
evaluated stochastically by the importance sampling, 
instead of preparing the Gutzwiller wave function
itself on a quantum computer. 
In general, this approach suffers from
the sign problem (more precisely, the phase problem)
as in the auxiliary-field quantum Monte Carlo method~\cite{Assaad_Evertz_book,Becca_Sorella_book}.

\subsection{Linear combination of unitary operators}\label{sec:method1}

\begin{figure}
  \begin{center}
    \includegraphics[width=.95\columnwidth]{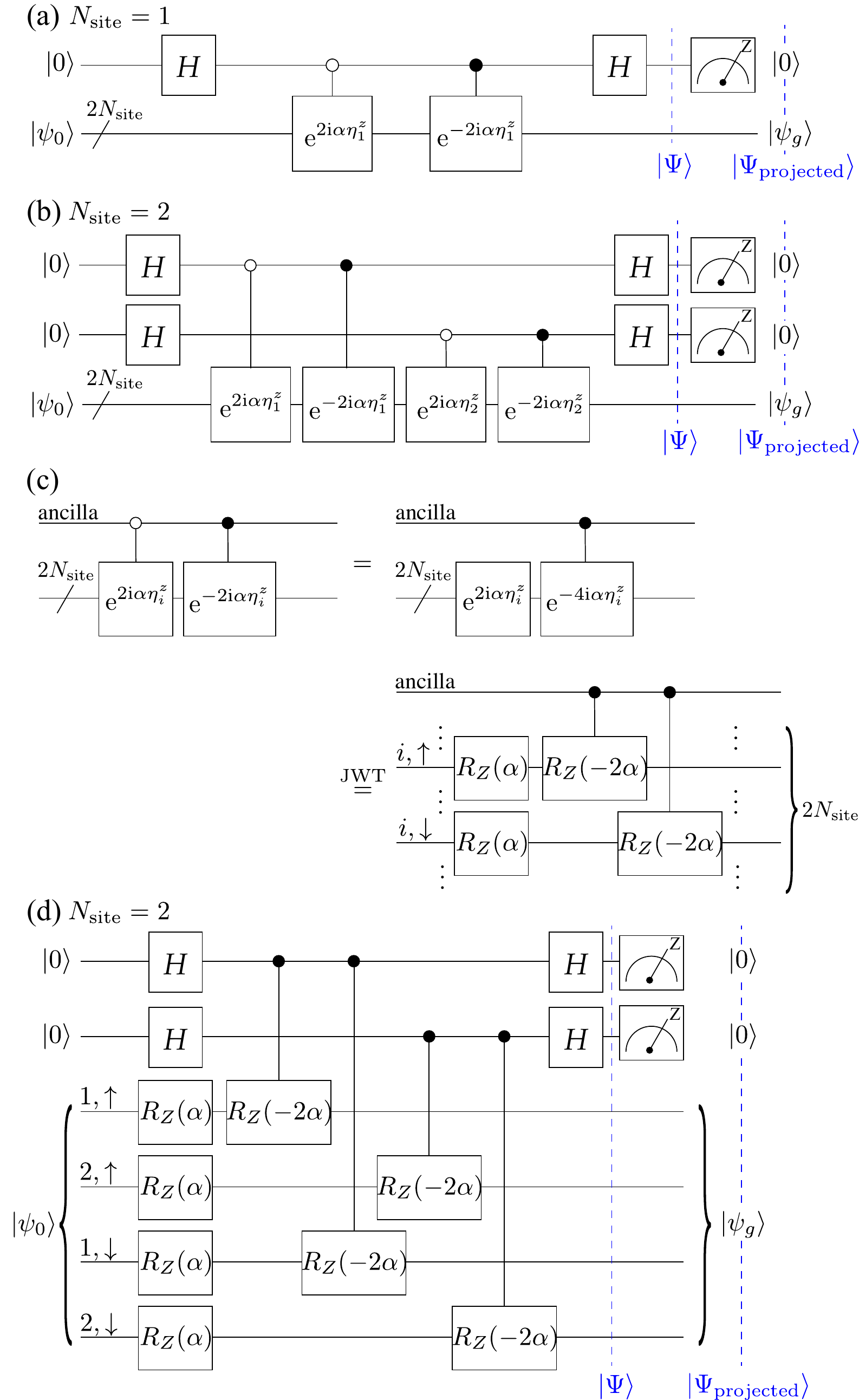}\\
    \caption{
      \label{fig:LCU}
      Quantum circuits for generating
      the Gutzwiller wave function $|\psi_g\rangle$ with
      $N_{\rm site}$ ancillary qubits and 
      $2N_{\rm site}$ register qubits 
      for 
      (a) $N_{\rm site}=1$ and
      (b) $N_{\rm site}=2$.
      $H$ denotes the Hadamard gate,  
      $|\Psi\rangle$ denotes the state of the whole (3$N_{\rm site}$-qubit) system
      after applying the second Hadamard gates on ancillary qubits, and
      $|\Psi_{\rm projected}\rangle$ denotes the state of the whole system
      after observing that the state of the ancillary qubits is $|00\cdots 0\rangle$.
      (c) Simplification of 
      a product of the two consecutive controlled-$\hat{\eta}^z_i$-rotation gates with opposite rotation angles, 
      assuming the Jordan-Wigner transformation in Eq.~(\ref{eq:etarot}).
      (d) The same as (b) but
      the controlled-$\hat{\eta}_i^z$ gates are now explicitly written with the simplification strategy 
      in (c). 
    }
  \end{center}
\end{figure}

As shown in Eq.~(\ref{eq:HST2}), summing all the auxiliary fields $s_i=\pm1$ for $i=1,2,\dots,N_{\rm site}$, 
the nonunitary Gutzwiller factor $\e^{-g\hat{D}}$ is expressed as a linear combination of $2^{N_{\rm site}}$ unitary operators, 
each of which is composed of a product of $N_{\rm site}$ unitary operators $\e^{\pm2\imag\alpha\hat{\eta}_i^z}$. 
Under the Jordan-Wigner transformation, the unitary operator $\e^{\pm2\imag\alpha\hat{\eta}_i^z}$ is then simply represented as 
a direct product of two single-qubit $Z$ rotation gates acting on qubits $i_\uparrow$ and $i_\downarrow$, as shown in Eq.~(\ref{eq:etarot}). 

In order to implement the linear combination of these unitary operators on a quantum computer, 
we can use a Hadamard-test-like variant~\cite{seki2021} of the quantum circuit known as 
the linear combination of unitary operators~\cite{Childs2012,kosugi2019construction},   
shown in Figs.~\ref{fig:LCU}(a) and \ref{fig:LCU}(b) for $N_{\rm site}=1$ and $2$, respectively, 
as examples, which can be easily generalized to $N_{\rm site}\geqslant3$. 
In these figures, we use $N_{\rm sites}$ ancillary qubits and measure each of them once. 
Instead, we can also consider the equivalent quantum circuit with only one ancillary qubit 
and every time after measuring it, we reuse this ancillary qubit repeatedly $N_{\rm site}$ times. 
This can certainly reduce the total number of necessary qubits, but has to initialize a qubit during the computation. 

Figure~\ref{fig:LCU}(c) shows another way to simplify the quantum circuits in Figs.~\ref{fig:LCU}(a) and \ref{fig:LCU}(b). 
Namely, the consecutive opposite-conditional
controlled-$\hat{\eta}_i^z$-rotation gates with 
opposite rotation angles
can be simplified by, for example, 
removing the ``controlled'' part from
the first controlled-$\hat{\eta}_i^z$-rotation gate and doubling the rotation angle
in the second controlled-$\hat{\eta}_i^z$-rotation gate.
Such an operation can be implemented, under the Jordan-Wigner transformation, with
2 $R_Z$ gates and 2 controlled-$R_Z$ gates, instead of 4 controlled-$R_Z$ gates, 
as shown in the lower part of Fig.~\ref{fig:LCU}(c). 
Following this strategy, the quantum circuit shown in Fig.~\ref{fig:LCU}(b) for $N_{\rm site}=2$ is now explicitly 
given in Fig.~\ref{fig:LCU}(d). 
We note that this simplification strategy is applicable
not only for a quantum circuit containing two consecutive controlled-$\hat{\eta}_i^z$-rotation gates with opposite rotation angles 
but also for a quantum circuit containing two consecutive controlled-time-evolution operators with opposite evolution times, 
and hence the quantum circuit proposed for the quantum power method in Ref.~\cite{seki2021} 
(and also a recent proposal for performing the imaginary-time evolution 
in Ref.~\cite{kosugi2021probabilistic}) can be simplified in the same manner.

If the measured states in the $N_{\rm site}$ ancillary qubits 
are all found in the state $|0\rangle$, then
the desirable state
$\prod_{i=1}^{N_{\rm site}}\sum_{s_i=\pm 1}\e^{2\imag \alpha s_i \hat{\eta}_i^z} |\psi_0\rangle$
is prepared in the rest of the qubits, i.e., in the $2N_{\rm site}$ register qubits, as shown in Fig.~\ref{fig:LCU}(d). 
After applying the second Hadamard gates on the ancillary qubits,
the state $|\Psi\rangle$ of the whole system (see Fig.~\ref{fig:LCU}) is given as
\begin{alignat}{1}
  |\Psi\rangle &=
  |00\cdots0\rangle \otimes
  \frac{1}{2^{N_{\rm site}}}
  \prod_{i=1}^{N_{\rm site}}
  ( \e^{2 \imag \alpha \hat{\eta}_i^z}
  + \e^{-2 \imag \alpha \hat{\eta}_i^z})
  |\psi_0\rangle \notag \\
  &+{\textrm{(unwanted terms)}} \notag \\
  &=|00\cdots0\rangle \otimes
  \e^{-gN_{\rm site}/4} \e^{-g\hat{D}}
  |\psi_0\rangle \notag \\
  &+{\textrm{(unwanted terms)}}.
  \label{eq:Psi}
\end{alignat}
Here, $|00\cdots 0\rangle$ denotes
the product state of all the $N_{\rm site}$ ancillary states being $|0\rangle$,   
and ``${\textrm{(unwanted terms)}}$'' denotes the other 
$2^{N_{\rm sites}}-1$ terms with the $N_{\rm site}$ ancillary states being 
distinct from $|00\cdots 0\rangle$, for which the Gutzwiller wave function $|\psi_g\rangle$ 
is not prepared in the register qubits.  
Note also that Eq.~(\ref{eq:HST2}) is used in the second equality of Eq.~(\ref{eq:Psi}).
According to the Born rule, the probability for successfully 
preparing the desired state $|\psi_g\rangle$, denoted as $p_{00\cdots0}$, 
is given by 
\begin{alignat}{1}
  p_{00 \cdots 0}
   = \langle \Psi |
   \hat{\cal P}_{00\cdots0}
  |\Psi \rangle 
   = \e^{-gN_{\rm site}/2}
  \langle \psi_0 | \e^{-2g\hat{D}} |\psi_0 \rangle, 
  \label{eq:p000}
\end{alignat}
where $\hat{\cal P}_{00\cdots0}=
|00\cdots0\rangle \langle 00\cdots 0|
\otimes \hat{I} $
is the projection operator that projects a state
in the whole Hilbert space onto the subspace
associated with the result of the measurement
observing that the state of the ancillary qubits is $|00\cdots0\rangle$.
According to the projection postulate, 
the state after the corresponding (successful) measurement,
denoted as $|\Psi_{\rm projected}\rangle$, is then given by 
\begin{equation}
  |\Psi_{\rm projected}\rangle=
  \frac{1}{\sqrt{p_{00\cdots0}}} \hat{\cal P}_{00\cdots 0}|\Psi \rangle
       =|00\cdots 0\rangle \otimes |\psi_g\rangle,
\end{equation}
indicating that the Gutzwiller wave function $|\psi_g\rangle$ is
prepared in the resister qubits. 

\begin{figure}
  \begin{center}
    \includegraphics[width=.95\columnwidth]{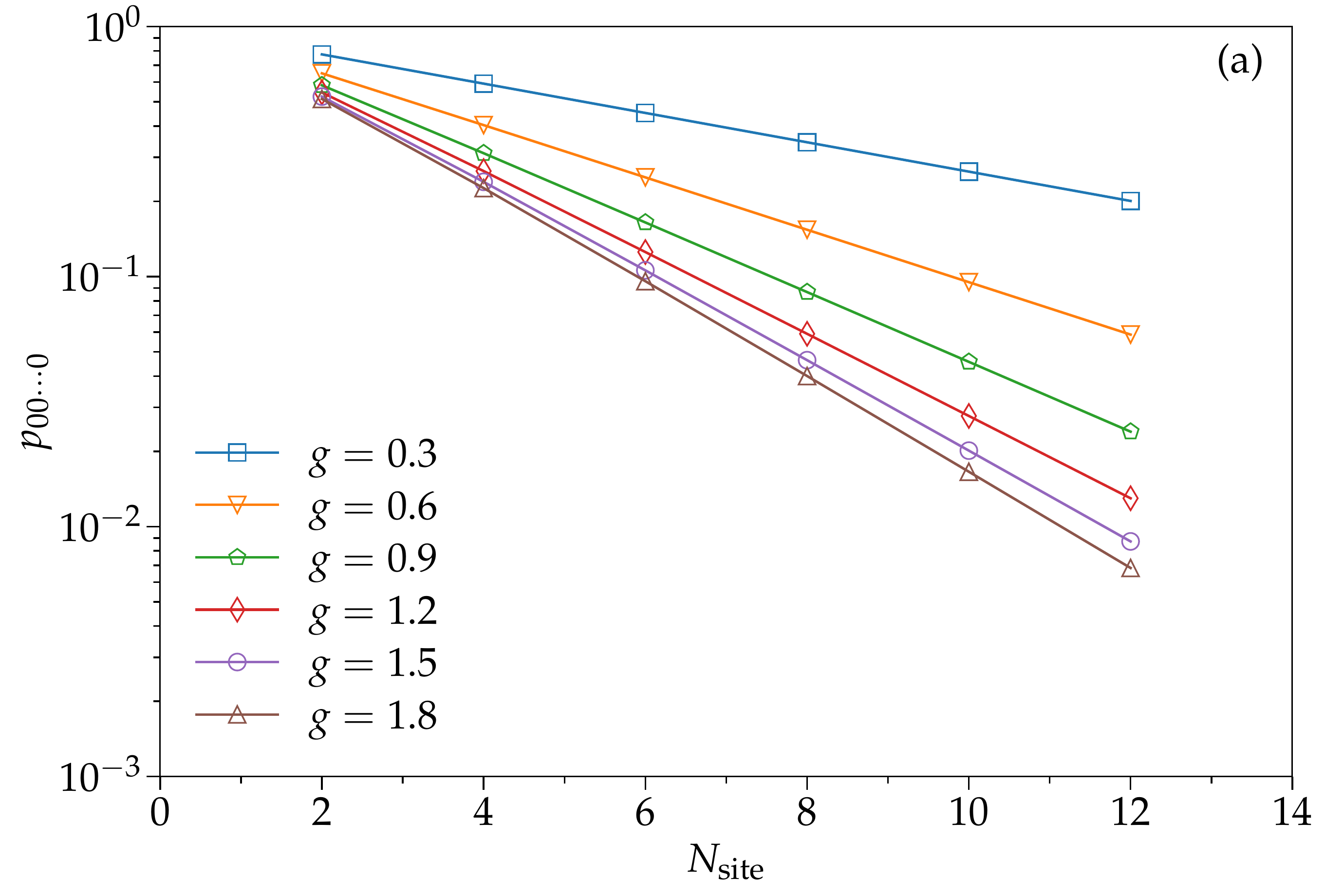}\\
    \includegraphics[width=.95\columnwidth]{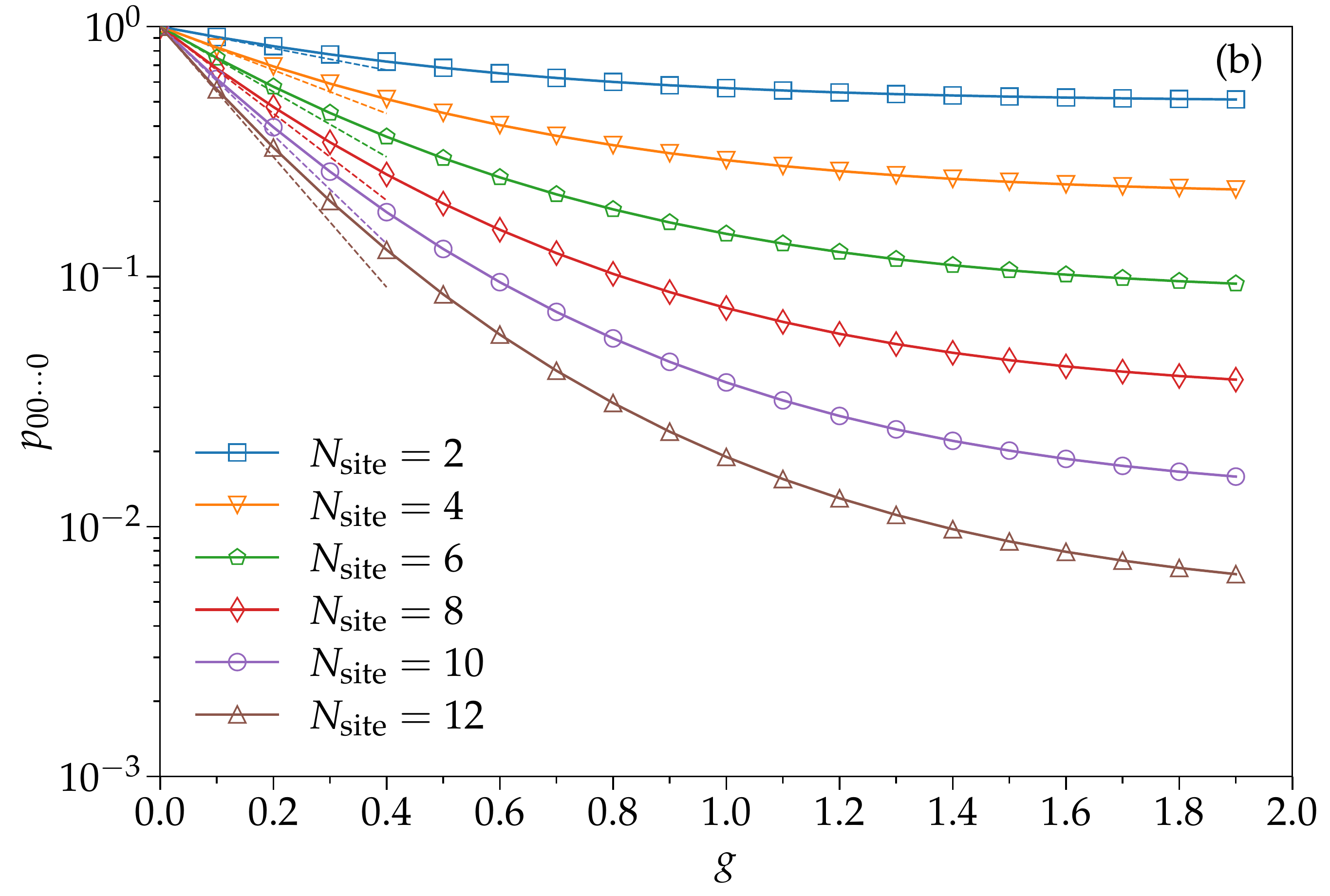}
    \caption{
      \label{fig:success}
      The success probability $p_{00\cdots0}$
      (a) as a function of $N_{\rm site}$ for several values of $g$ and 
      (b) as a function of $g$ for several values of $N_{\rm site}$. 
      The dashed lines in (b) indicate the exponential
      decrease of $p_{00\cdots0}$ for small $g$ according 
      to Eq.~(\ref{eq:initslope}). 
      The trial state $|\psi_0\rangle$
      is chosen as the ground state of $\hat{K}$
      at half filling 
      defined on a one-dimensional chain under open-boundary conditions. 
      Solid lines are guide for the eyes.
    }
  \end{center}
\end{figure}

Figure~\ref{fig:success} shows the success probability 
$p_{00\cdots0}$ as functions of $N_{\rm site}$ and $g$ 
calculated numerically using a classical computer.
Here, the trial state $|\psi_0\rangle$ is chosen as the ground state of $\hat{K}$ 
at half filling 
defined on the one-dimensional chain under open-boundary conditions. 
As clearly observed in Fig.~\ref{fig:success}(a),
the success probability $p_{00\cdots0}$ decrease exponentially in $N_{\rm site}$.
In order to examine the $g$ dependence of $p_{00\cdots0}$, it is useful to 
study the logarithmic derivative of the success probability.  
It follows from Eq.~(\ref{eq:p000}) that 
the logarithmic derivative of the success probability, $\partial_g \ln p_{00\cdots0}$, 
is related to the expectation value of $\hat{D}$ via
\begin{alignat}{1}
  \langle \psi_g|\hat{D}|\psi_g \rangle
  &= -\frac{1}{2} \frac{\partial}{\partial g}
  \ln \langle \psi_0|\e^{-2g\hat{D}}|\psi_0\rangle \notag \\
  &= -\left(\frac{N_{\rm site}}{4}+
  \frac{1}{2}
  \frac{\partial}{\partial g}\ln p_{00\cdots0}\right).
  \label{eq:expecD}
\end{alignat}
Since
$\lim_{g\to 0} \langle \psi_g|\hat{D}|\psi_g \rangle=0$ and 
$\lim_{g\to \infty} \langle \psi_g|\hat{D}|\psi_g \rangle = -N_{\rm site}/4$ 
for the present choice of $|\psi_0\rangle$,
the slopes of $\ln p_{00\cdots0}$ in the two limits are given respectively by 
\begin{equation}
  \lim_{g\to0} \frac{\partial}{\partial g}\ln{p_{00\cdots0}} = -\frac{N_{\rm site}}{2}
  \label{eq:initslope}
\end{equation}
and
\begin{equation}
  \lim_{g\to\infty} \frac{\partial}{\partial g}\ln{p_{00\cdots0}} = 0,
  \label{eq:endslope}
\end{equation}
implying that $p_{00\cdots0}$ decreases 
exponentially in $g$ for small $g$, but 
the decrease saturates for large $g$, as indeed found 
in Fig.~\ref{fig:success}(b).
A finite success probability in the limit $g\to \infty$
leaves a possibility of preparing
the Gutzwiller-projected state~\cite{Gros1987} relevant for the 
$t$-$J$-type models~\cite{Harris1967,Hirsch1985,Zhang1988,Eskes1994,Eskes1996,Otsuka2002,Eder2011} 
for a moderate $N_{\rm site}$.  

Despite the exponential decrease of the success probability in $N_{\rm site}$,
an advantage of the approach described here is that the Gutzwiller wave function $|\psi_g\rangle$
itself can be prepared on a quantum computer.
In this sense,
the present scheme is similar to that in the previous study~\cite{Murta2021}, 
even though the two schemes take different routes: 
the quantum circuit in Ref.~\cite{Murta2021} is based on the original form of the Gutzwiller factor in Eq.~(\ref{eq:G})
with a slightly different parametrization, whereas
the quantum circuit in the present study is based on the Hubbard-Stratonovich-transformed
Gutzwiller factor in Eq.~(\ref{eq:HST2}).
It should be emphasized that the quantum circuit proposed here 
is rather simpler than that proposed in the previous study~\cite{Murta2021}.
This point will be further discussed in Sec.~\ref{sec:conclusions}.

\subsection{Importance sampling}\label{sec:method2}

\subsubsection{Reformulation and sampling}

To perform the sum $\sum_{\bs{s}}$ over the auxiliary fields $\bs{s}=\{\{s_{i,\tau}\}_{i=1}^{N_{\rm site}}\}_{\tau=1}^2$ in Eq.~(\ref{eq:O}) 
stochastically using the Monte Carlo method,
we rewrite Eq.~(\ref{eq:O}) as
\begin{alignat}{1}
  \langle \hat{O}\rangle
  &=
  \sum_{\bs{s}}
  P_{\bs{s}}
  \langle \hat{O} \rangle_{\bs{s}},
  \label{eq:IS}
\end{alignat}
where 
\begin{equation}
  P_{\bs{s}}
  \equiv
  \frac{
    \langle\psi_0 |
    \prod_{i=1}^{N_{\rm site}}            
    \e^{2 \imag \alpha (s_{i,1}+s_{i,2}) \hat{\eta}_i^z}
    |\psi_0 \rangle
  }
       {
         \sum_{\bs{s}^\prime}
         \langle\psi_0 |
         \prod_{i=1}^{N_{\rm site}}            
         \e^{2 \imag \alpha (s_{i,1}^\prime+s_{i,2}^\prime) \hat{\eta}_i^z}
         |\psi_0 \rangle}
         \label{eq:ps}
\end{equation}
and 
\begin{equation}
  \langle \hat{O}\rangle_{\bs{s}}
  \equiv \frac
  {
    \langle\psi_0 |
    \prod_{i=1}^{N_{\rm site}}            
    \e^{2 \imag \alpha s_{i,2} \hat{\eta}_i^z}
    \hat{O}
    \prod_{j=1}^{N_{\rm site}}            
    \e^{2 \imag \alpha s_{j,1} \hat{\eta}_i^z}
    |\psi_0 \rangle
  }
  {
    \langle\psi_0 |
    \prod_{i=1}^{N_{\rm site}}            
    \e^{2 \imag \alpha (s_{i,1}+s_{i,2}) \hat{\eta}_i^z}
    |\psi_0 \rangle}.
  \label{eq:Os}
\end{equation}
Notice that
$P_{\bs{s}}$ is in general complex
and hence the method suffers from the phase problem, 
as in the auxiliary-field Monte Carlo method~\cite{Assaad_Evertz_book,Becca_Sorella_book}.
In the presence of the phase problem,
a proper modification in Eq.~(\ref{eq:IS}) 
is necessary (see for example Refs.~\cite{Imada1989,Loh1990,Hamann1990}).
In this study, however, we only consider cases
satisfying that $P_{\bs{s}}$ is real and $P_{\bs{s}}>0$, i.e., in the absence of the phase problem (see Appendix~\ref{sec:pp}). 

The auxiliary fields in Eq.~(\ref{eq:IS}) are sampled
by the Metropolis-Hastings algorithm
using the local update with an acceptance probability 
$p(\bs{s}\to \bs{s^\prime})=\min (1,P_{\bs{s}^\prime}/P_{\bs{s}})$
for accepting the move from $\bs{s}$ to $\bs{s}^\prime$. 
In the local update, the candidate auxiliary fields $\bs{s}^\prime=\{\{s'_{i,\tau}\}_{i=1}^{N_{\rm site}}\}_{\tau=1}^2$ 
is chosen by flipping only a single auxiliary field, $s_{i,\tau} \to -s_{i,\tau}$, 
among the current auxiliary fields $\bs{s}=\{\{s_{i,\tau}\}_{i=1}^{N_{\rm site}}\}_{\tau=1}^2$ 
and the remaining auxiliary fields are unaltered. 
If the proposed move from $\bs{s}$ to $\bs{s}^\prime$ is accepted, the candidate auxiliary fields $\bs{s}'$ 
are adopted as the new auxiliary fields for the next iteration. 
Otherwise, the old auxiliary fields $\bs{s}$ remain for the next iteration. 
Here, we select a flipped auxiliary field $s_{i,\tau}$ in the candidate auxiliary fields $\bs{s}'$ 
sequentially for $i=1,2,\dots,N_{\rm site}$ and $\tau=1,2$, 
and define one Monte Carlo sweep when all the auxiliary fields are selected once in the Monte Carlo iterations. 
We measure observables every Monte Carlo sweep and denote the number of measurements by $N_{\rm MC}$. 
Note that we do not have to evaluate the denominator in Eq.~(\ref{eq:ps}) because only the ratio of 
$P_{\bs{s}^\prime}/P_{\bs{s}}$ is required in the Monte Carlo iterations.

\begin{figure}
  \begin{center}
    \includegraphics[width=.95\columnwidth]{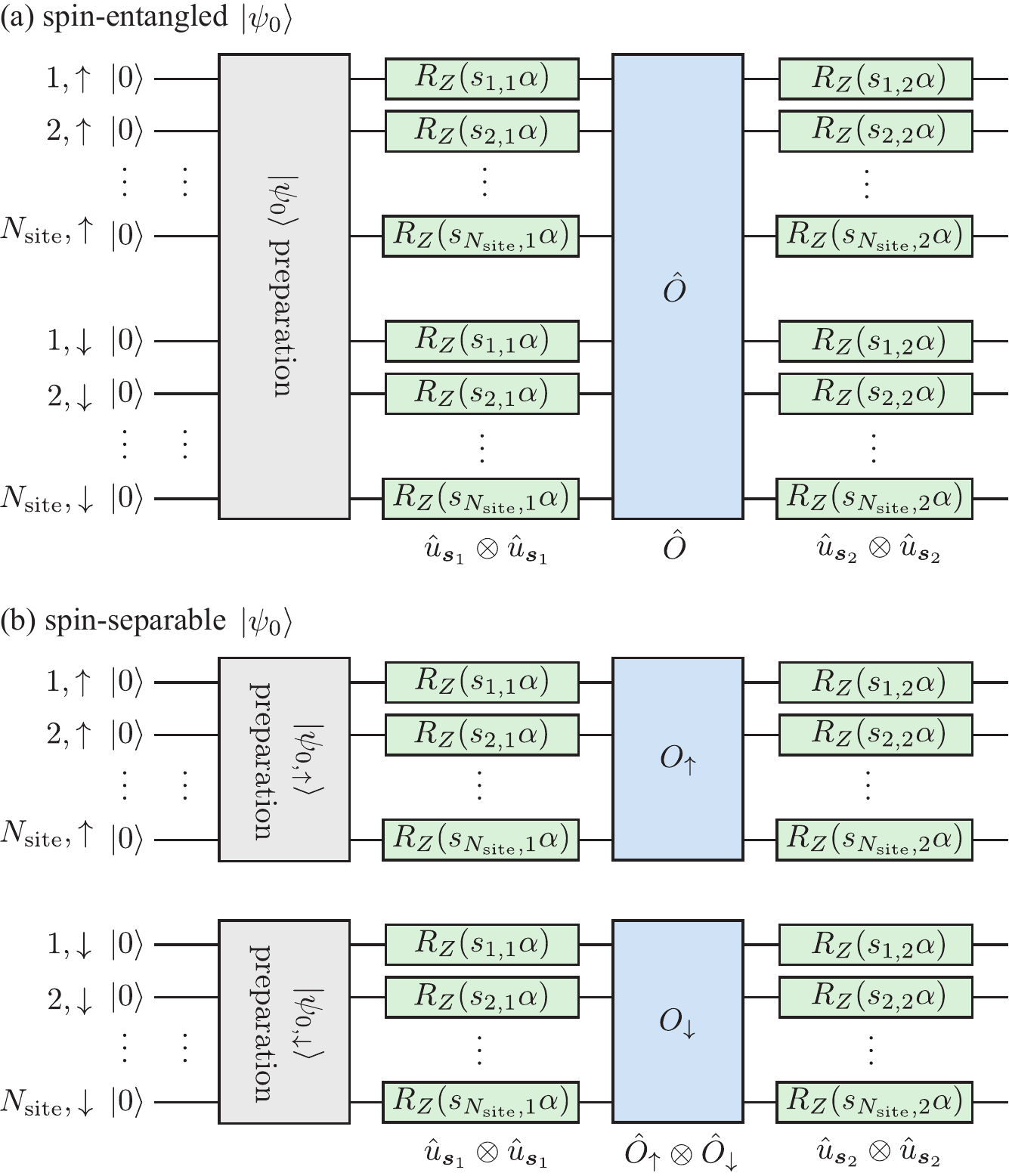}
    \caption{
      \label{fig:circuits}
      Quantum circuits for preparing the
      state $
      \prod_{i}\e^{2\imag \alpha s_{i,2} \hat{\eta}_i^z}
      \hat{O}
      \prod_{i}\e^{2\imag \alpha s_{i,1} \hat{\eta}_i^z}
      |\psi_0 \rangle$ 
      appearing in the numerator of Eq.~(\ref{eq:Os})
      for (a) a spin-entangled state $|\psi_0\rangle$ and
      (b) a spin-separable state $|\psi_0\rangle$, 
      assuming the Jordan-Wigner transformation. 
      We also assume that the observable $\hat{O}$ is spin-separable in (b). 
    }
  \end{center}
\end{figure}

Using the relation in Eq.~(\ref{eq:etarot}) for $\hat{\eta}_i$ under the Jordan-Wigner transformation, 
the numerator of $P_{\bs{s}}$ in Eq.~(\ref{eq:ps}) and the denominator of $\langle \hat{O} \rangle_{\bs{s}}$ 
in Eq.~(\ref{eq:Os}) can be expressed with the Pauli operators as 
\begin{alignat}{1}
&\langle\psi_0 |
\prod_{i=1}^{N_{\rm site}}            
\e^{2 \imag \alpha (s_{i,1}+s_{i,2}) \hat{\eta}_i^z}
|\psi_0 \rangle \nonumber \\
\overset{\rm JWT}{=}&
\langle\psi_0 |
\prod_{i=1}^{N_{\rm site}}
\hat{R}_{Z_{i_\uparrow}}\left((s_{i,1}+s_{i,2})\alpha\right)  \hat{R}_{Z_{i_\downarrow}}\left((s_{i,1}+s_{i,2})\alpha\right)
|\psi_0 \rangle, 
\end{alignat}
where the symbol ``$\otimes$" for a direct product is omitted for simplicity. 
Similarly, the numerator of $\langle \hat{O} \rangle_{\bs{s}}$ 
in Eq.~(\ref{eq:Os}) can be given as 
\begin{alignat}{1}
&\langle\psi_0 |
    \prod_{i=1}^{N_{\rm site}}            
    \e^{2 \imag \alpha s_{i,2} \hat{\eta}_i^z}
    \hat{O}
    \prod_{j=1}^{N_{\rm site}}            
    \e^{2 \imag \alpha s_{j,1} \hat{\eta}_i^z}
    |\psi_0 \rangle \nonumber \\
\overset{\rm JWT}{=}&
\langle\psi_0 |
    \prod_{i=1}^{N_{\rm site}}            
    \hat{R}_{Z_{i_\uparrow}}(s_{i,2}\alpha)  \hat{R}_{Z_{i_\downarrow}}(s_{i,2}\alpha)
    \hat{O}
    \prod_{j=1}^{N_{\rm site}}            
    \hat{R}_{Z_{j_\uparrow}}(s_{j,1}\alpha)  \hat{R}_{Z_{j_\downarrow}}(s_{j,1}\alpha)
    |\psi_0 \rangle. 
\end{alignat}
Therefore, as shown in Fig.~\ref{fig:circuits}(a), a quantum circuit for preparing
the state 
$
\prod_{i}\e^{2\imag \alpha s_{i,2} \hat{\eta}_i^z}
\hat{O}
\prod_{i}\e^{2\imag \alpha s_{i,1} \hat{\eta}_i^z}
|\psi_0 \rangle$
in the numerator of $\langle \hat{O} \rangle_{\bs{s}}$ in Eq.~(\ref{eq:Os}) is significantly simple. 
A quantum circuit for the state $\prod_{i=1}^{N_{\rm site}}            
    \e^{2 \imag \alpha (s_{i,1}+s_{i,2}) \hat{\eta}_i^z}
    |\psi_0 \rangle$ 
in the denominator of $\langle \hat{O} \rangle_{\bs{s}}$ in Eq.~(\ref{eq:Os}) can
be obtained by simply setting $\hat{O}=\hat{I}$ and combining the two rotation gates 
$\hat{R}_{Z_{i_\sigma}}(s_{i,1}\alpha)$ and $\hat{R}_{Z_{i_\sigma}}(s_{i,2}\alpha)$
into the single rotation $\hat{R}_{Z_{i_\sigma}}((s_{i,1}+s_{i,2})\alpha)$ in Fig.~\ref{fig:circuits}(a).

\subsubsection{Simplification for spin-separable states}\label{subsec:spin_separable}

Next, we describe a simplification that can be 
applied when the trial state $|\psi_0\rangle$ is separable according to the
decomposition of the Hilbert space 
${\cal V}={\cal V}_{\uparrow} \otimes {\cal V}_{\downarrow}$,
where ${\cal V}_{\sigma}$ denotes the Hilbert space for fermions with spin $\sigma$. 
Let us assume that $|\psi_0\rangle$ is a separable state of the form 
\begin{equation}
  |\psi_0\rangle =
  |\psi_{0,\uparrow}\rangle \otimes
  |\psi_{0,\downarrow} \rangle,
  \label{eq:separable}
\end{equation}
where $|\psi_{0,\sigma}\rangle \in {\cal V}_{\sigma}$, 
as in the case for the ground state of $\hat{K}$. 
We refer to a state of the form in Eq.~(\ref{eq:separable}) as a spin-separable state.

Let us now introduce the following unitary operator 
\begin{equation}
  \hat{u}_{\bs{s}_\tau,{\sigma}} \equiv \prod_{i=1}^{N_{\rm site}}
  \e^{\imag \alpha s_{i,\tau} (\hat{n}_{i\sigma}-\frac{1}{2})}
  \overset{\rm JWT}{=}
  \prod_{i=1}^{N_{\rm site}}\hat{R}_{Z_{i_\sigma}}\left(s_{i,\tau} \alpha \right) 
  \label{eq:us}
\end{equation}
on ${\cal V}_\sigma$ for a given set of auxiliary fields $\bs{s}_{i,\tau}=\{s_{i,\tau}\}_{i=1}^{N_{\rm site}}$. 
The last equality is simply because 
$
\hat{n}_{i\sigma} \overset{\rm JWT}{=} \frac{1}{2}(1 - \hat{Z}_{i_\sigma})
$
under the Jordan-Wigner transformation. 
Then the product of unitary operators generated by $\hat{\eta}^z_i$ 
can be written as 
\begin{equation}
  \prod_{i=1}^{N_{\rm site}} \e^{2\imag \alpha (s_{i,1}+s_{i,2})\hat{\eta}_{i}}
  =\hat{u}_{\bs{s}_2,{\uparrow}}\hat{u}_{\bs{s}_1,{\uparrow}} \otimes
   \hat{u}_{\bs{s}_2,{\downarrow}}\hat{u}_{\bs{s}_1,{\downarrow}}.  
\end{equation}
Therefore, $P_{\bs{s}}$ and
$\langle \hat{O} \rangle_{\bs{s}}$ in Eqs.~(\ref{eq:ps}) and (\ref{eq:Os}) 
are given respectively as 
\begin{equation}
  P_{\bs{s}}
  =
  \frac{
  \langle \psi_{0,\uparrow}   | \hat{u}_{\bs{s}_2,{\uparrow}} \hat{u}_{\bs{s}_1,{\uparrow}} | \psi_{0,\uparrow}\rangle
  \langle \psi_{0,\downarrow} | \hat{u}_{\bs{s}_2,{\downarrow}} \hat{u}_{\bs{s}_1,{\downarrow}} | \psi_{0,\downarrow}\rangle
  }{\sum_{\bs{s}^\prime}
   \langle \psi_{0,\uparrow}  | \hat{u}_{\bs{s}_2^\prime,{\uparrow}} \hat{u}_{\bs{s}_1^\prime,{\uparrow}} | \psi_{0,\uparrow}\rangle
  \langle \psi_{0,\downarrow} | \hat{u}_{\bs{s}_2^\prime,{\downarrow}} \hat{u}_{\bs{s}_1^\prime,{\downarrow}} | \psi_{0,\downarrow}\rangle}
  \label{eq:ps2}
\end{equation}
and 
\begin{equation}
  \langle \hat{O} \rangle_{\bs{s}} =
  \frac{
  \langle \psi_{0,\uparrow}   | \hat{u}_{\bs{s}_2,{\uparrow}} \hat{O}_\uparrow   \hat{u}_{\bs{s}_1,{\uparrow}}| \psi_{0,\uparrow}\rangle
}{
  \langle \psi_{0,\uparrow}   | \hat{u}_{\bs{s}_2,{\uparrow}} \hat{u}_{\bs{s}_1,{\uparrow}} | \psi_{0,\uparrow}\rangle
  }
  \cdot
  \frac{
  \langle \psi_{0,\downarrow} | \hat{u}_{\bs{s}_2,{\downarrow}} \hat{O}_\downarrow \hat{u}_{\bs{s}_1,{\downarrow}}| \psi_{0,\downarrow}\rangle
}{
  \langle \psi_{0,\downarrow} | \hat{u}_{\bs{s}_2,{\downarrow}} \hat{u}_{\bs{s}_1,{\downarrow}} | \psi_{0,\downarrow}\rangle},
  \label{eq:Os2}
\end{equation}
where the observable of the form 
\begin{equation}
  \hat{O} = \hat{O}_{\uparrow} \otimes \hat{O}_{\downarrow} 
\label{eq:separable_O}
\end{equation}
is assumed. 
A quantum circuit for preparing
the state 
$
\prod_{i}\e^{2\imag \alpha s_{i,2} \hat{\eta}_i^z}
\hat{O}
\prod_{i}\e^{2\imag \alpha s_{i,1} \hat{\eta}_i^z}
|\psi_0 \rangle$
in the numerator of $\langle \hat{O} \rangle_{\bs{s}}$ in Eq.~(\ref{eq:Os}) 
is now further simplified as shown in Fig.~\ref{fig:circuits}(b).
A quantum circuit for the state $\prod_{i=1}^{N_{\rm site}}            
    \e^{2 \imag \alpha (s_{i,1}+s_{i,2}) \hat{\eta}_i^z}
    |\psi_0 \rangle$
in the denominator of $\langle \hat{O} \rangle_{\bs{s}}$ in Eq.~(\ref{eq:Os}) can
be obtained by simply setting $\hat{O}=\hat{I}$ and combining two consecutive rotations 
into one in Fig.~\ref{fig:circuits}(b).
Notice that 
when $|\psi_0\rangle$ is a spin-separable state,
only $N_{\rm site}$ qubits are required at a time.

Finally, we note that since the kinetic term of the Hamiltonian has the form, 
\begin{equation}
  \hat{K} = \hat{K}_{\uparrow} \otimes \hat{I} + \hat{I} \otimes \hat{K}_{\downarrow},  
\end{equation}
where $\hat{K}_{\sigma}$ is the summand of $\sum_{\sigma}$ in Eq.~(\ref{eq:K}), 
$\langle \hat{K} \rangle_{\bs{s}} $ can be written simply as 
\begin{equation}
  \langle \hat{K} \rangle_{\bs{s}} =
  \frac{
    \langle \psi_{0,\uparrow} | \hat{u}_{\bs{s}_2} \hat{K}_\uparrow \hat{u}_{\bs{s}_1}| \psi_{0,\uparrow}\rangle
  }{
    \langle \psi_{0,\uparrow} | \hat{u}_{\bs{s}_2} \hat{u}_{\bs{s}_1} | \psi_{0,\uparrow}\rangle}
  +
  \frac{
  \langle \psi_{0,\downarrow} | \hat{u}_{\bs{s}_2} \hat{K}_\downarrow \hat{u}_{\bs{s}_1}| \psi_{0,\downarrow}\rangle
}{
  \langle \psi_{0,\downarrow} | \hat{u}_{\bs{s}_2} \hat{u}_{\bs{s}_1} | \psi_{0,\downarrow}\rangle}.
\end{equation}
A similar formula is also obtained for $\langle \hat{D} \rangle_{\bs{s}}$.

\subsubsection{Numerical simulations}

To demonstrate the method proposed here, we employ a classical computer to evaluate numerical  
the expectation values of the total energy $E= \langle \hat{K} \rangle + U \langle \hat{D} \rangle$,
the kinetic energy $\langle \hat{K} \rangle$, and
the potential energy $U \langle \hat{D} \rangle$
as a function of $g$ for several $U$ values. 
The trial state $|\psi_0\rangle$ is chosen as
the ground state of $\hat{K}$, which is a spin-separable state, and in this case the Monte Carlo 
importance sampling can be performed without the phase problem. 
Although we use a classical computer to demonstrate the method proposed here, 
we briefly comment on how to prepare the trial state $|\psi_0\rangle$ on a quantum computer. 
Generally, a Slater-determinant state can be
prepared on a quantum computer with at most $O(N_{\rm site}^2)$ number of two-qubit (e.g., Givens)
rotation gates starting from a relevant product state~\cite{Wecker2015,Kivlichan2018,Jiang2018}.
A concrete example of variationally preparing the ground state of $\hat{K}$ in one dimension
using a discretized quantum-adiabatic process can be found in Ref.~\cite{Shirakawa2021}.

\begin{figure*}
  \begin{center}
    \includegraphics[width=1.00\textwidth]{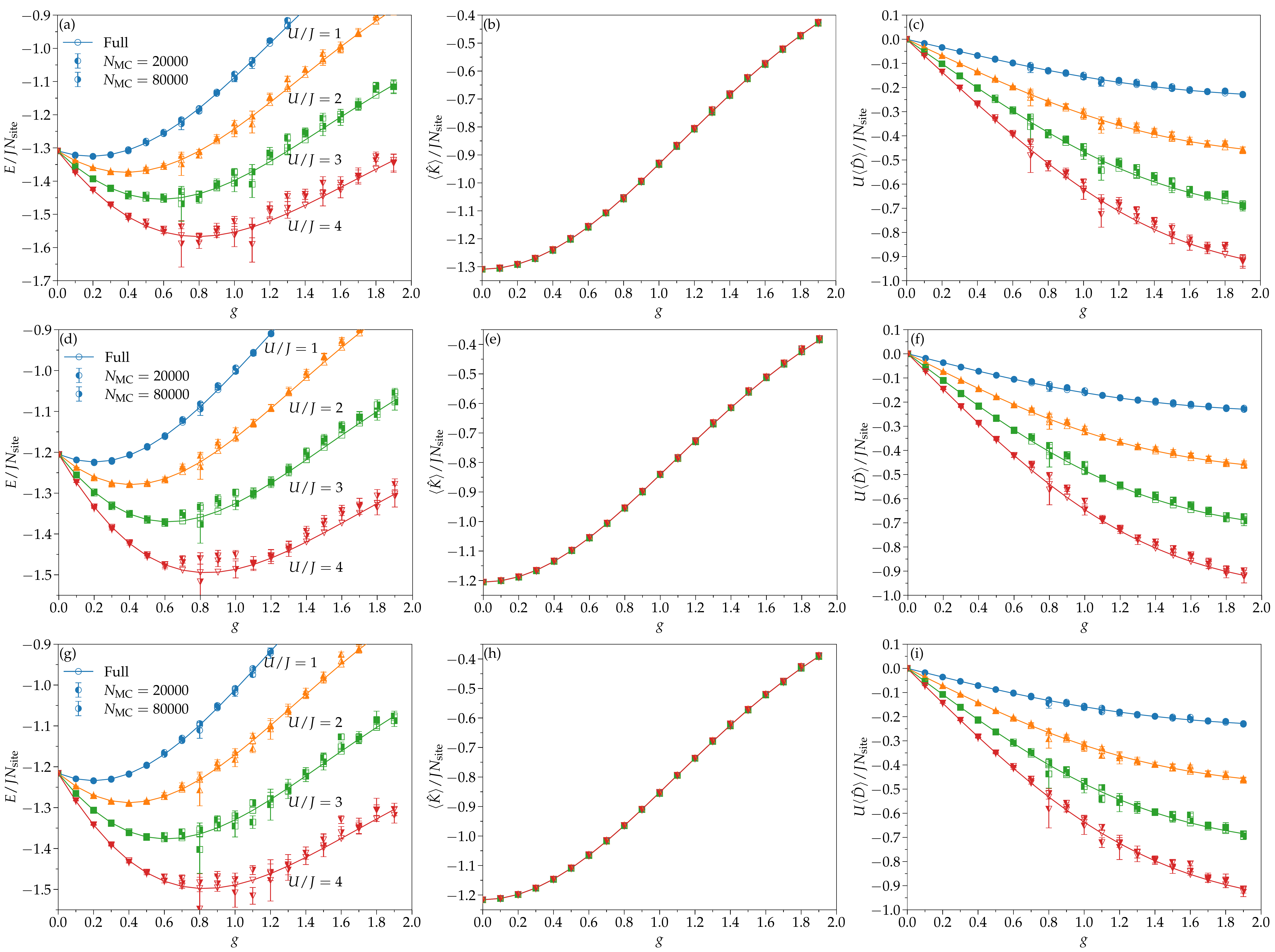}
    \caption{
      \label{fig:energyU4}
      The expectation values of (a, d, g) the total energy $E$,
      (b, e, h) the kinetic energy $\langle \hat{K}\rangle$, and
      (c, f, i) the potential energy $U\langle \hat{D} \rangle$ 
      as a function of the Gutzwiller parameter $g$ for the Fermi-Hubbard model 
      at half filling with $U/J=1,2,3,4$ (top to bottom).
      The models studied here are on (a-c) a $4\times 2$ ladder lattice, (d-f) a 10-site one-dimensional 
      lattice, and (g-i) a 12-site one-dimensional lattice under open-boundary conditions. 
      The number $N_{\rm MC}$ of Monte Carlo samplings is indicated in the figure. 
      For comparison, the exact results are also plotted by solid lines. In addition, the results obtained by fully summing all the 
      auxiliary fields are shown by open symbols. 
      Note that $\langle \hat{K}\rangle$ does not depends on the value of $U$. The results of $\langle \hat{K}\rangle$ 
      for two different $N_{\rm MC}$ values and obtained by fully summing all the 
      auxiliary fields are indistinguishable in this scale. 
    }
  \end{center}
\end{figure*}

Figure~\ref{fig:energyU4} shows the results for Fermi-Hubbard model in three different lattices with 
$N_{\rm MC}=20000$ and $N_{\rm MC}=80000$. 
For comparison, the exact results as well as the results obtained by fully summing all the auxiliary fields are 
also shown. Although the proposed method can reproduce the exact results within the statistical errors, we find that    
the total energy $E$ has the larger statistical errors for 
larger $U/J$ in all the three lattices. 
By resolving the total energy $E$ into
the kinetic and the potential energies, we find that
the kinetic energy has the much smaller statistical errors than 
the potential energy.
Therefore, the large statistical errors in $E$ are mainly due to the
large statistical errors in the potential energy, i.e., the expectation value of $\hat{D}$.
We note that the larger statistical errors in the potential
energy are due to the choice of the auxiliary fields, which are coupled to
the local fermion-density operator $\hat{\eta}_i^z$:
if the auxiliary fields are coupled to the local spin operator~\cite{Hirsch1983} 
(although in this case the Gutzwiller factor is no longer expressed as
a linear combination of unitary operators and hence this is less relevant
in the context of this study), 
the fluctuation of $\langle \hat{D} \rangle$ can be suppressed.

\section{Demonstration on a quantum device}\label{sec:results}

In this section, we shall use a NISQ device to demonstrate the proposed method for
the two-site Fermi-Hubbard model. 
As described below, the Gutzwiller wave function 
can describe the exact ground state of the two-site Fermi-Hubbard model.

\subsection{Gutzwiller wave function for the two-site Fermi-Hubbard model at half filling}
First, we review the well-known fact that 
the ground state of the two-site Fermi-Hubbard model at half
filling can be described by the Gutzwiller wave function with
$|\psi_0\rangle$ being the ground state of $\hat{K}$ 
(for example, see Ref.~\cite{Fabrizio2007}). 
Let $|{\rm vac}\rangle$ 
be the fermion vacuum such that $\hat{c}_{i\sigma}|{\rm vac}\rangle=0$
for any site $i$ and spin $\sigma$. 
Then the ground state of $\hat{K}$ at half filling
is given by the following state with two fermions occupying the bonding orbital:  
\begin{alignat}{1}
  |\psi_0\rangle
  =& \frac{1}{2}
  \left(\hat{c}_{1\uparrow}^\dag+\hat{c}_{2\uparrow}^\dag\right)
  \left(\hat{c}_{1\downarrow}^\dag+\hat{c}_{2\downarrow}^\dag\right)
  |{\rm vac}\rangle.
  \label{eq:p02}
\end{alignat}
A straightforward calculation shows that 
$\langle \psi_0 | \e^{-g\hat{D}} \hat{K} \e^{-g\hat{D}}|\psi_0\rangle=-2J$,
$\langle \psi_0 | \e^{-2g\hat{D}}|\psi_0\rangle=\cosh{g}$, and
$\langle \psi_0 | \e^{-g\hat{D}} \hat{D} \e^{-g\hat{D}}|\psi_0\rangle=
-\frac{1}{2}\partial_g \langle \psi_0 | \e^{-2g\hat{D}}|\psi_0\rangle
= -\frac{1}{2}\sinh{g}$. 
Therefore, the total energy $E(g)$ is given by 
\begin{alignat}{1}
  E(g)
  &= \langle \hat{K} \rangle + U \langle \hat{D} \rangle
  \notag \\
  &=-\frac{1}{\cosh{g}}\left(2J+\frac{U}{2}\sinh{g} \right).
  \label{eq:Eg}
\end{alignat}

Considering $E(g)$ as the variational energy, 
the stationary condition $\partial E(g)/\partial g|_{g=g_{\rm opt}}=0$ gives us 
the optimal variational parameter $g_{\rm opt}$ such that 
\begin{equation}
  \sinh{g_{\rm opt}} = \frac{U}{4J},
  \label{eq:sinhg}
\end{equation}
or equivalently 
$g_{\rm opt}=\ln\left(\frac{U}{4J}+\sqrt{\left(\frac{U}{4J}\right)^2+1}\right)$. 
By substituting Eq.~(\ref{eq:sinhg}) into
Eq.~(\ref{eq:Eg}), the optimized variational energy is obtained as 
\begin{equation}
  E(g_{\rm opt}) = -\sqrt{4J^2 + \frac{U^2}{4}}, 
\end{equation}
which coincides with the exact ground-state energy of
the two-site Fermi-Hubbard model at half filling, implying that the Gutzwiller wave function $|\psi_g\rangle$ with $g=g_{\rm opt}$ 
is the exact ground state of the two-site Fermi-Hubbard model at half filling.

\subsection{Quantum simulations} 

Using a real quantum device, 
we shall now evaluate the expectation values of the total energy $E$, the kinetic energy $\langle\hat{K}\rangle$, and the 
potential energy $U\langle\hat{D}\rangle$ with respect to the Gutzwiller wave function $|\psi_g\rangle$ 
for the two-site Fermi-Hubbard model at half filling. 
Under the Jordan-Wigner transformation,
the fermion vacuum is expressed as $|{\rm vac}\rangle
\overset{\rm JWT}{=}
|0\rangle_{1_\uparrow}
|0\rangle_{2_\uparrow}
|0\rangle_{1_\downarrow}
|0\rangle_{2_\downarrow}$
and hence $|\psi_0\rangle$ in Eq.~(\ref{eq:p02}), i.e., the ground state of $\hat{K}$, 
is given by 
$
|\psi_{0} \rangle 
=
| \psi_{0, \uparrow} \rangle \otimes  | \psi_{0, \downarrow} \rangle
$ with
\begin{align}
	| \psi_{0,\sigma} \rangle 
	=&
	\frac{1}{\sqrt{2}}
	\left(
	  |0\rangle_{1_\sigma}|1\rangle_{2_\sigma}
	+ |1\rangle_{1_\sigma}|0\rangle_{2_\sigma}
	\right).
\label{eq:Bell}
\end{align}
This state $| \psi_{0,\sigma} \rangle$ is merely one of the Bell states and is easily prepared as
\begin{align}
| \psi_{0,\sigma} \rangle 
=
\widehat{CX} \left( \hat{H} \otimes \hat{X} \right)
|0\rangle_{1_\sigma} |0\rangle_{2_\sigma},
\end{align}
where $\hat{X}$, $\hat{H}$ and $\widehat{CX}$ denote the gate operations for 
Pauli $X$, Hadamard, and controlled-$X$ (CNOT) gates, respectively.
As discussed in Sec.~\ref{subsec:spin_separable},
for the spin-separable trial state $|\psi_0\rangle$, 
the numerator of Eq.~(\ref{eq:O}) is expressed as
\begin{equation}
 \langle \psi_{0} | \e^{-g \hat{D}} 
\hat{O}_{\uparrow} \otimes \hat{O}_{\downarrow} 
\e^{-g \hat{D}} | \psi_{0} \rangle
 =
 \gamma^4
 \sum_{\bs{s}}
 \prod_{\sigma=\uparrow, \downarrow}
 P_{\bs{s} \sigma}^{O_{1}O_{2}}(g)
\end{equation}
with 
\begin{equation}
 P_{\bs{s} \sigma}^{O_{1}O_{2}}(g) \equiv
 \langle \psi_{0,\sigma } |
 \hat{u}_{\bs{s}_2,\sigma}
 \hat{O}_{1_\sigma}  \hat{O}_{2_\sigma}
 \hat{u}_{\bs{s}_1,\sigma}
 | \psi_{0,\sigma } \rangle,
 \label{eq:OO}
\end{equation}
where $\hat{O}_{\sigma} = \hat{O}_{1_\sigma} \hat{O}_{2_\sigma}$ 
and $\hat{O}_{i_\sigma}=\{\hat{I}_{i_\sigma},\hat{X}_{i_\sigma},\hat{Y}_{i_\sigma},\hat{Z}_{i_\sigma}\}$.
Here, $\hat{I}_{i_\sigma}$ is the identity operator acting on the $i_\sigma$ qubit. 
Noticing that $\hat{u}_{\bs{s}_1,\sigma} \neq \hat{u}_{\bs{s}_2,\sigma}$ in general,
we can not use the direct measurement method to evaluate the expectation value 
of Eq.~(\ref{eq:OO}), at least in a simple way~\cite{Mitarai_PRR2018}. 
Alternatively,
we exploit the Hadamard test to measure the expectation value of the unitary operator
$\hat{u}_{\bs{s}_2,\sigma} \hat{O}_{\sigma} \hat{u}_{\bs{s}_1,\sigma}$~\cite{Tacchino_AQT2020}.
The specific form of the Hamiltonian for the two-site Fermi-Hubbard model 
is simply given by
$\hat{\cal H} = \hat{K} + U\hat{D}$ with
$
 \hat{K} \overset{\rm JWT}{=} - \frac{J}{2} \sum_{\sigma=\uparrow, \downarrow}
\left( 
 \hat{X}_{1_\sigma  } \hat{X}_{2_\sigma  }
 + \hat{Y}_{1_\sigma  } \hat{Y}_{2_\sigma  }
 \right)
$ and
$
 \hat{D} \overset{\rm JWT}{=}   \frac{1}{4} \sum_{i=1, 2}
 \hat{Z}_{i_\uparrow  } \hat{Z}_{i_\downarrow}
$.
Therefore, it is sufficient to evaluate
$P_{\bs{s} \uparrow}^{II}(g)$,
$P_{\bs{s} \uparrow}^{ZI}(g)$,
and
$P_{\bs{s} \uparrow}^{XX}(g)$
to calculate $\langle \hat{K} \rangle$ and $\langle \hat{D} \rangle$,
and thereby the total energy $E$.
Instead of employing the importance sampling,
here we directly perform the sum over all the auxiliary fields,
$
\sum_{\bs{s}} \cdots = 
\sum_{s_{1,1}=\pm1}
\sum_{s_{2,1}=\pm1}
\sum_{s_{1,2}=\pm1}
\sum_{s_{2,2}=\pm1} \cdots
$,
because the total number of terms is only $2^4=16$.

\begin{figure*}[tb]
 \includegraphics[height=60pt]{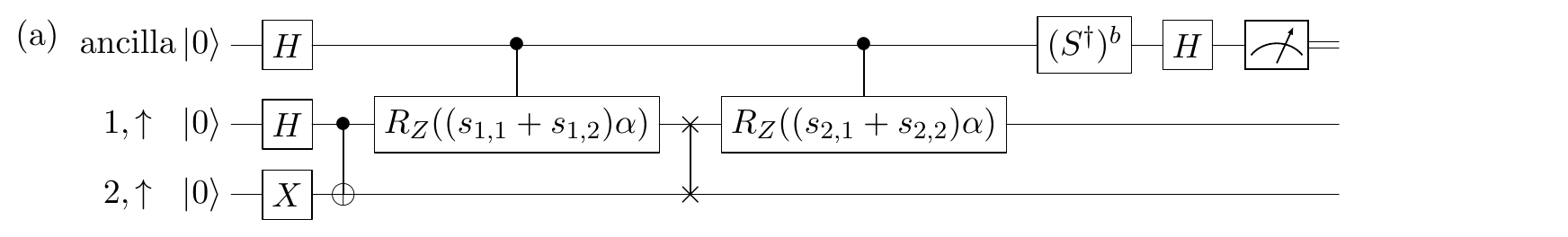}
 \includegraphics[height=60pt]{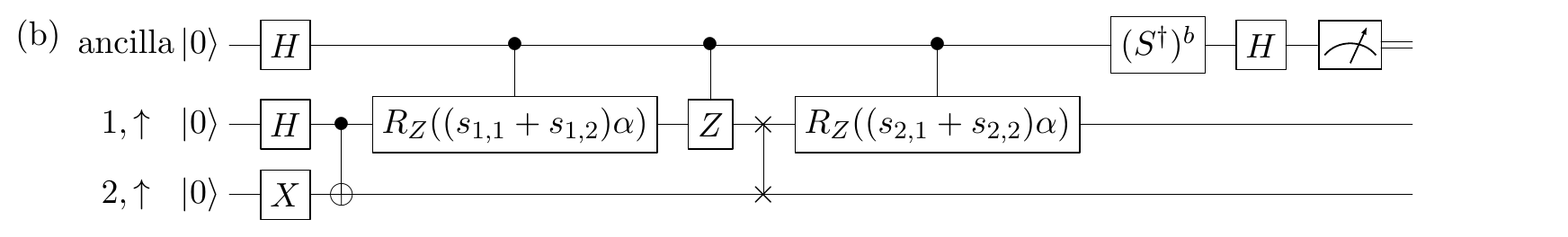}
 \includegraphics[height=60pt]{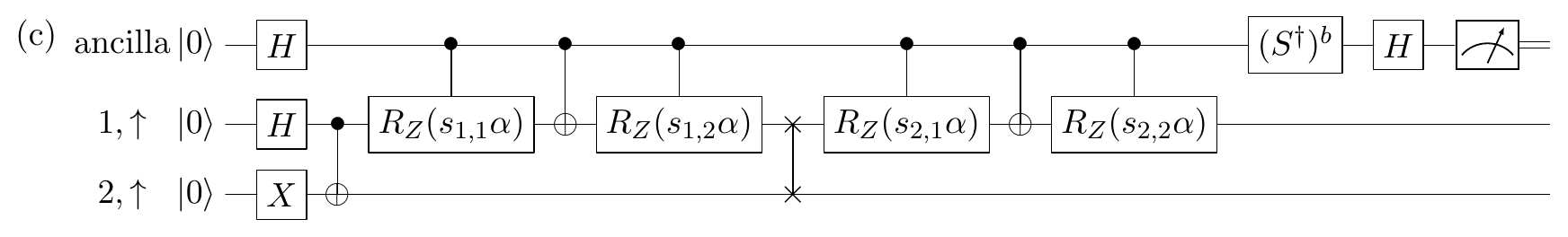}
 \caption{
 \label{fig:Hadamard}
 Quantum circuits for the Hadamard test to evaluate 
 the expectation values of 
 (a) 
 $
 P_{\bs{s} \uparrow}^{II}(g)=
 \langle \psi_{0\uparrow} |  \hat{u}_{\bs{s}_2,\uparrow} 
 \hat{u}_{\bs{s}_1,\uparrow} | \psi_{0\uparrow} \rangle$,
 (b) 
 $
 P_{\bs{s} \uparrow}^{ZI}(g)=
\langle \psi_{0\uparrow} |  \hat{u}_{\bs{s}_2,\uparrow} 
 \hat{Z}_{1_\uparrow} 
 \hat{u}_{\bs{s}_1,\uparrow} | \psi_{0\uparrow} \rangle$,
 and
 (c) 
 $
 P_{\bs{s} \uparrow}^{XX}(g)=
 \langle \psi_{0\uparrow} |  \hat{u}_{\bs{s}_2,\uparrow} 
 \hat{X}_{1_\uparrow} \hat{X}_{2_\uparrow}
 \hat{u}_{\bs{s}_1,\uparrow} | \psi_{0\uparrow} \rangle$.
 The real and imaginary part of each expectation value are estimated as 
 $P_{0} - P_{1}$ for $b=0$ and $1$, respectively.
 Here, $P_{0}$ is the probability of measuring $|0\rangle$ 
 at the ancillary qubit and $P_{1}=1-P_{0}$.
 The SWAP gate is denoted by a line connecting two crosses. 
 $S^\dag$ denotes the single-qubit phase shift gate acting as $\hat{S}^\dag|0\rangle_{i_\sigma} = |0\rangle_{i_\sigma}$ and 
 $\hat{S}^\dag|1\rangle_{i_\sigma} = -\imag |1\rangle_{i_\sigma}$.
 }
\end{figure*}

Figure~\ref{fig:Hadamard} shows the quantum circuits for estimating the expectation values 
$P_{\bs{s} \uparrow}^{II}(g)$,
$P_{\bs{s} \uparrow}^{ZI}(g)$,
and
$P_{\bs{s} \uparrow}^{XX}(g)$. 
Notice that here we explicitly introduce the SWAP operations in order to 
involve only two-qubit gates acting on neighboring qubits in these quantum circuits. 
To further simplify the quantum circuits, 
the second SWAP gate that would have primarily 
been placed after the last $Z$ rotation in each quantum circuit is omitted 
without affecting validity of the measurement. 
This latter simplification, yielding reduction of three CNOT gates,
is beneficial for suppressing noise inherent to a real quantum device.

We implement the quantum circuits using the Quantum Information 
Software Kit (Qiskit)~\cite{Qiskit} and perform computations on 
the IBM Q Manila device (\texttt{ibmq\_manila}), 
the device publicly available 
through the IBM Quantum Lab platform~\cite{IBM2}.
We also run the same quantum circuits 
on the classical simulator (\texttt{qasm\_simulator}),
which is considered as an ideal quantum device, 
to realize the impact of noise.
Each experiment runs 8192 shots to measure the local state at the ancillary qubit in the computational basis.  
The same set of experiments is repeated 16 times to evaluate the average and the standard deviation, 
the latter being the estimate of the statistical error.

\begin{figure*}[tb]
 \centering
 \includegraphics[width=.95\textwidth]{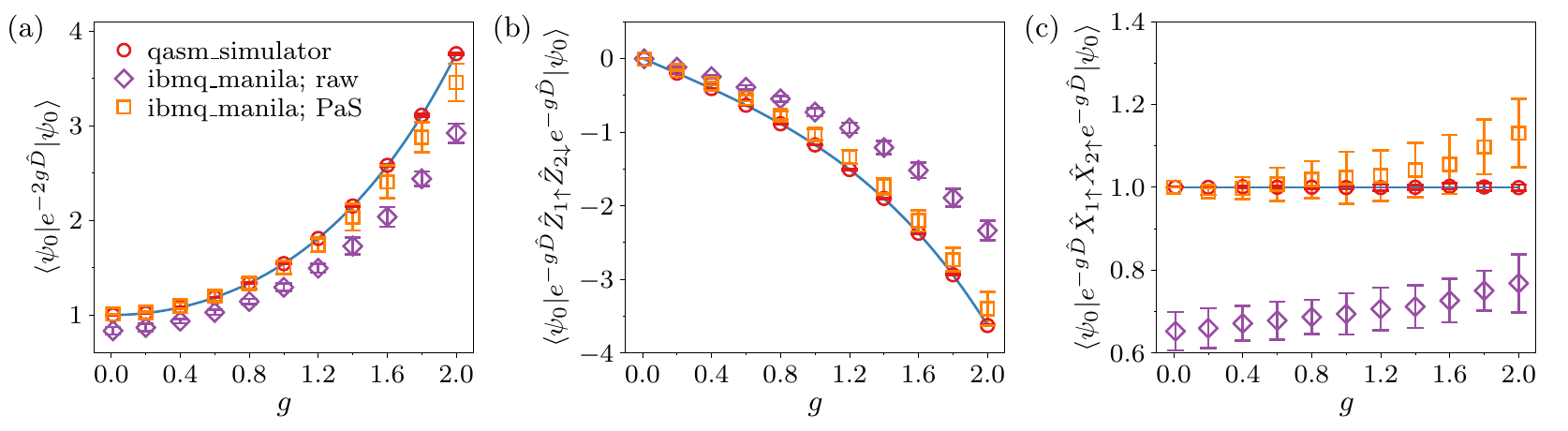}
 \caption{
 \label{fig:nisq-OOOO}
 Results of 
 (a) $\langle \psi_{0} | \e^{-2g\hat{D}} | \psi_{0} \rangle$,
 (b) $\langle \psi_{0} | \e^{-g\hat{D}} \hat{Z}_{1_\uparrow} \hat{Z}_{1_\downarrow} \e^{-g\hat{D}} | \psi_{0} \rangle$, and
 (c) $\langle \psi_{0} | \e^{-g\hat{D}} \hat{X}_{1_\uparrow} \hat{X}_{2_\uparrow  } \e^{-g\hat{D}} | \psi_{0} \rangle$
 as a function of the Gutzwiller parameter $g$.
 The results evaluated from the raw data obtained from the IBM Q Manila device are denoted by diamonds, and
 those after the error mitigation by the PaS correction technique are shown by squares. 
 For comparison, the results calculated on the classical simulator (\texttt{qasm\_simulator}) using the same quantum circuits
 are also plotted by circles, which agree with the analytical results shown by solid lines.
 }
\end{figure*}

Figure~\ref{fig:nisq-OOOO} shows the results for
the denominator of Eq.~(\ref{eq:O}) and 
the numerators for
$\langle \hat{Z}_{1_\uparrow} \hat{Z}_{1_\downarrow} \rangle$ and
$\langle \hat{X}_{1_\uparrow} \hat{X}_{2_\uparrow} \rangle$,
which are calculated respectively as
\begin{alignat}{1}
  &\langle \psi_{0} | \e^{-2g\hat{D}} | \psi_{0} \rangle 
  = \left( \sum_{\bs{s}} P_{\bs{s} \uparrow}^{II}(g) \right)^{2},
  \\
  &\langle \psi_{0} | \e^{-g\hat{D}} \hat{Z}_{1_\uparrow} \hat{Z}_{1_\downarrow} \e^{-g\hat{D}} | \psi_{0} \rangle 
  = \left( \sum_{\bs{s}} P_{\bs{s} \uparrow}^{ZI}(g) \right)^{2},
\end{alignat}
and 
\begin{alignat}{1}
\langle \psi_{0} | \e^{-g\hat{D}} \hat{X}_{1_\uparrow} \hat{X}_{2_\uparrow} \e^{-g\hat{D}} | \psi_{0} \rangle 
 = \left( \sum_{\bs{s}} P_{\bs{s} \uparrow}^{XX}(g) \right)
    \left( \sum_{\bs{s}} P_{\bs{s} \uparrow}^{II}(g) \right).
\end{alignat}
Here, we utilize the equivalence between fermion spins $\uparrow$ and $\downarrow$.
Despite the simple quantum circuits,
we notice the sizable discrepancies between the results computed directly from 
the real quantum device 
and the analytical results.
The discrepancies are more noticeable for 
$\langle \psi_{0} | \e^{-g\hat{D}} \hat{X}_{1_\uparrow} \hat{X}_{2_\uparrow} \e^{-g\hat{D}} | \psi_{0} \rangle$,
as shown in Fig.~\ref{fig:nisq-OOOO}(c),
since the evaluation of $P_{\bs{s} \uparrow}^{XX}(g)$ involves more CNOT gates 
than the others (see Fig.~\ref{fig:Hadamard}). 

\begin{figure*}[tb]
 \centering
 \includegraphics[width=.95\textwidth]{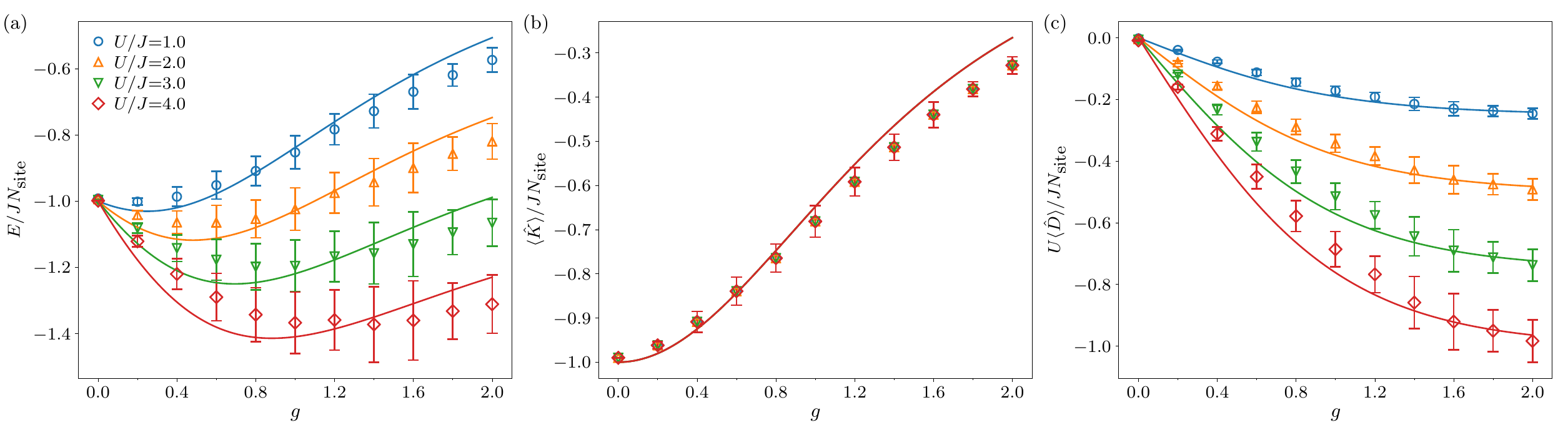}
 \caption{
 \label{fig:nisq-EKD}
 The expectation values of (a) the total energy $E$,
 (b) the kinetic energy $\langle \hat{K}\rangle$, and
 (c) the potential energy $U\langle \hat{D} \rangle$ 
  as a function of the Gutzwiller parameter $g$ for the two-site ($N_{\rm site}=2$) Fermi-Hubbard model 
  under open-boundary conditions at half filling with $U/J=1,2,3,4$ (top to bottom).
 The results are obtained from 
 the error-mitigated values shown in Fig.~\ref{fig:nisq-OOOO}. 
 For comparison, the exact analytical results are also plotted by solid lines. 
 }
\end{figure*}

To mitigate the systematic errors,
we apply a so-called phase-and-scale (PaS) correction technique
developed in the study of spin dynamics~\cite{Chiesa2019,Francis_PRB2019}. 
Since the ideal value of $P_{\bs{s} \uparrow}^{II}(g)$ at $g=0$ is know to be one for all $\bs{s}$, 
in the PaS correction method, the inverse of the raw data of $P_{\bs{s} \uparrow}^{II}(0)$
is multiplied to the raw data of $P_{\bs{s} \uparrow}^{II}(g)$ 
to mitigate the systematic errors. 
The same strategy is applied to $P_{\bs{s} \uparrow}^{XX}(g)$ 
because $|\psi_{0}\rangle$ is chosen as the ground state of $\hat{K}$, i.e., a spin singlet state.
On the other hand, for the error mitigation in $P_{\bs{s} \uparrow}^{ZI}(g)$, 
we use the raw data at a sufficiently large $g$, i.e., $g=10$,   
so as to reproduce the value in the strong coupling limit, where the rotation angle $\alpha$ is 
simply $\frac{\pi}{2}$. 
This simple technique is found to successfully mitigate the most of the 
systematic errors, as shown in Fig.~\ref{fig:nisq-OOOO}.
Finally, the total energy $E$ as well as the kinetic energy $\langle\hat{K}\rangle$ and 
the potential energy $U\langle\hat{D}\rangle$ calculated from 
these error-mitigated values is shown in Fig.~\ref{fig:nisq-EKD}. 
We find that these results are in good agreement with the analytical results 
within the error bars.
We note that the PaS correction technique can in principle
be applied for larger systems because the exact values of
$\langle \hat{K} \rangle$ with $g=0$ in the noninteracting limit and
$\langle \hat{D} \rangle$ with $g\gg 1$ in the atomic limit
can be evaluated efficiently with classical computers or even analytically.

%-------------------------------------------------------------------------------

\section{Conclusion and discussion}\label{sec:conclusions}

Based on the discrete Hubbard-Stratonovich transformation
of the Gutzwiller factor $\e^{-g\hat{D}}$, 
we have proposed a scheme to implement the Gutzwiller
wave function on a quantum computer and demonstrated 
it using numerical simulations and a real quantum device.
The crucial point is that 
the nonunitary Gutzwiller factor $\e^{-g\hat{D}}$ is expressed as a linear combination of
unitary operators by introducing the auxiliary fields of the discrete Hubbard-Stratonovich transformation. 
The sum over the auxiliary fields generates an exponentially large number of terms 
with respect to the system size $N_{\rm site}$, which is here treated by 
two complementary approaches, 
one employing a quantum circuit of the linear combination of unitaries to probabilistically 
prepare the Gutzwiller wave function on a quantum computer and the other 
using the importance sampling to stochastically evaluate the expectation values for 
the Gutzwiller wave function. 

The first approach performs 
the sum over all the auxiliary fields 
by measuring the state of the $N_{\rm site}$ ancillary qubits.
Although the success probability decreases exponentially in $N_{\rm site}$,
an advantage of this approach is that the Gutzwiller wave function $|\psi_g\rangle$
itself can be prepared on a quantum computer.
In this sense, this approach 
is similar to that reported in Ref.~\cite{Murta2021}. 
Indeed, using Eqs.~(\ref{eq:G}) and (\ref{eq:tilde_g}), 
one can show that the success probability $p_{00\cdots0}$ is exactly 
the same as that in Ref.~\cite{Murta2021}.
Interestingly, however, the circuit structure
is different from that in Ref.~\cite{Murta2021}, as summarized in Table~\ref{Table}.
While $N_{\rm site}$ controlled-controlled-$R_Y$ (CC$R_Y$) gates are
required for preparing 
the Gutzwiller wave function in the previous study~\cite{Murta2021}, 
assuming the Jordan-Wigner transformation, 
here $2N_{\rm site}$ C$R_Z$ gates are used for the same purpose. 
In terms of the number of CNOT gates, 
the implementation of a C$R_Z$ gate
is simpler than that of a CC$R_Y$ gate because
a C$R_Z$ gate can be decomposed into 2 CNOT gates and 2 $R_Z$ gates, while
a CC$R_Y$ gate can be decomposed into 2 Toffoli gates and 2 $R_{Y}$ gates~\cite{Barenco1995}.
The CNOT gate counts 
for decomposing 2 C$R_{Z}$ gates is thus 4, whereas that
for decomposing a single CC$R_Y$ gate is 12 
because a Toffoli gate requires 6 CNOT gates~\cite{NielsenChuang}. 
Therefore, in terms of the CNOT gate counts, 
the present scheme is beneficial 
for preparing the Gutzwiller wave function on a quantum computer.

\begin{table*}
  \caption{
    Comparison of the gate counts in two quantum circuits proposed 
    in Ref~\cite{Murta2021} and 
    the present study (see Fig.~\ref{fig:LCU})
    for probabilistically preparing 
    the Gutzwiller wave function on a quantum computer. 
    The Jordan-Wigner transformation is
    assumed in both schemes. 
  \label{Table}}
    \begin{tabular}{lll}
      \hline
      \hline
      & 
      Ref.~\cite{Murta2021} &
      This study \\
      \hline
      ``Unit gate'' for implementing the Gutzwiller factor &
      CC$R_Y$ &
      2C$R_Z$ $+$ 2$R_Z$      \\
      Number of CNOT gates required for decomposing of a single unit gate&
      $12$ &
      $4$\\
      Number of unit gates required for implementing the Gutzwiller factor &
      $N_{\rm site}$  &
      $N_{\rm site}$       \\
      Total number of CNOT gates required for implementing the Gutzwiller factor &
      $12 N_{\rm site}$ &
      $4 N_{\rm site}$\\
      \hline
      \hline
    \end{tabular}
\end{table*}

We should emphasize that this simplification of the quantum circuit is made possible
because our quantum circuit is inspired by the Hubbard-Stratonovich transformation.  
The Hubbard-Stratonovich transformation allows us to disentangle 
the two-body operator $\e^{-g
  (\hat{n}_{i\uparrow}-\frac{1}{2})
  (\hat{n}_{i\downarrow}-\frac{1}{2})}$
into a linear combination of one-body operators $\propto
\e^{ i\alpha (\hat{n}_{i\uparrow}+\hat{n}_{i\downarrow}-1)}+
\e^{-i\alpha (\hat{n}_{i\uparrow}+\hat{n}_{i\downarrow}-1)}$,
each of which is then represented simply
as a direct product of single-qubit rotations
$R_{Z_{i_\uparrow}}\otimes R_{Z_{i_\downarrow}}$
under the Jordan-Wigner transformation. 

\begin{figure}
  \begin{center}
    \includegraphics[width=.95\columnwidth]{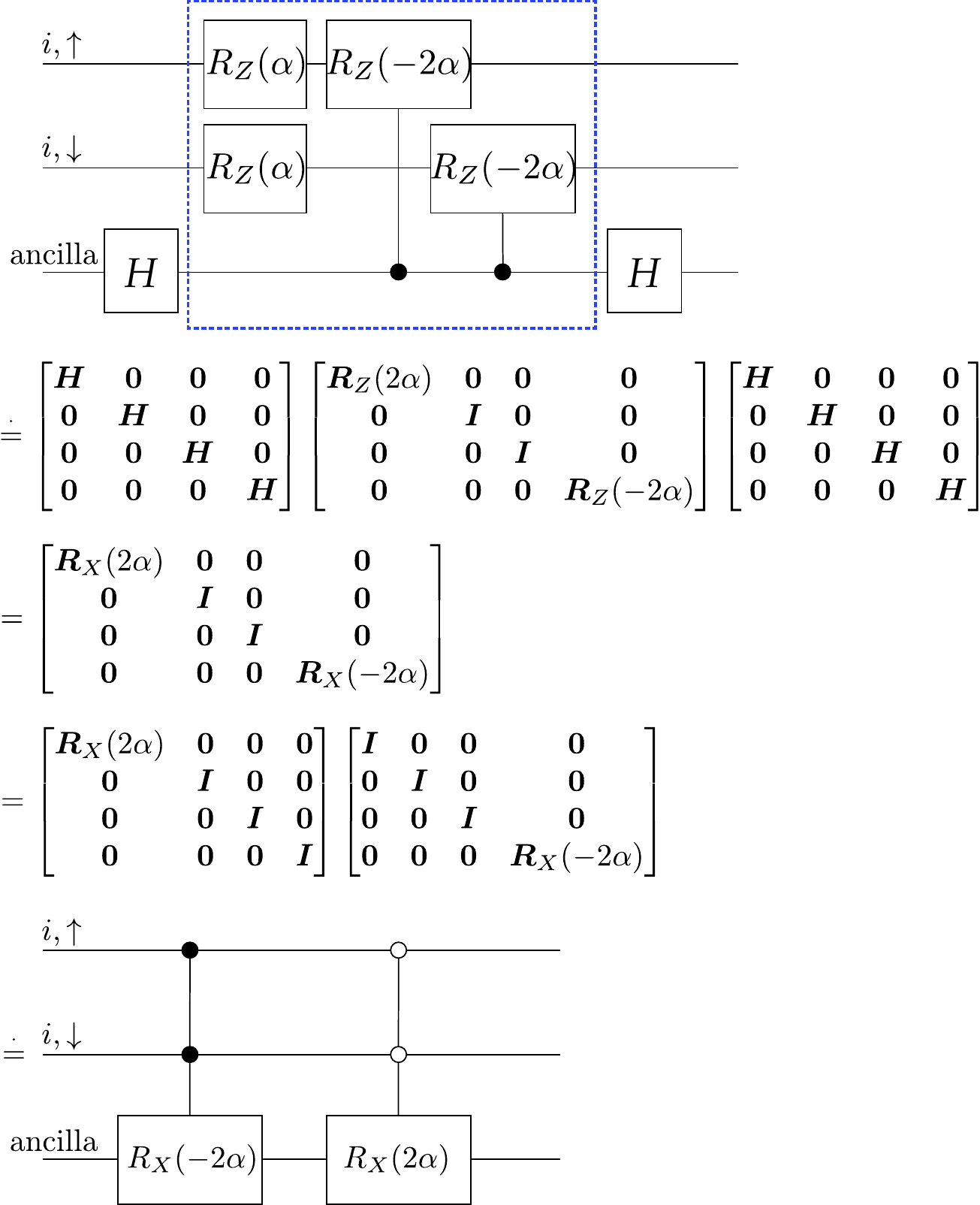}\\
    \caption{
      \label{fig:8x8}
      Different expression of 
      the quantum circuit for taking the linear combination of the one-body operators 
      $
      \frac{1}{2}(
      \e^{ i\alpha (\hat{n}_{i\uparrow}+\hat{n}_{i\downarrow}-1)}+
      \e^{-i\alpha (\hat{n}_{i\uparrow}+\hat{n}_{i\downarrow}-1)}
      )
      $
      with controlled-controlled-unitary gates.
      The quantum circuit in the first line is the same as that in the lower part of Fig.~\ref{fig:LCU}(c). 
      The matrices in the second line (third and fourth lines) 
      are $8\times 8$ matrices, corresponding to the quantum circuit in
      the first (last) line,
      with $\bs{0}$, $\bs{I}$, $\bs{H}$, $\bs{R}_Z$, and $\bs{R}_X$
      being the $2\times 2$ null, identity, Hadamard, $Z$-rotation, and $X$-rotation
      matrices, respectively.
      The diagonal matrix in the second line corresponds to the
      sequence of the gates enclosed by the blue dotted line in the first line. 
      The matrices are represented with respect to the basis states 
      $|0\rangle_{i_\uparrow}|0\rangle_{i_\downarrow}|0\rangle_{\rm ancilla}$,
      $|0\rangle_{i_\uparrow}|0\rangle_{i_\downarrow}|1\rangle_{\rm ancilla}$,
      $|0\rangle_{i_\uparrow}|1\rangle_{i_\downarrow}|0\rangle_{\rm ancilla}$,
      $\cdots$,
      $|1\rangle_{i_\uparrow}|1\rangle_{i_\downarrow}|1\rangle_{\rm ancilla}$. 
    }
    \end{center}
\end{figure}

Note, however, that 
the quantum circuit for taking the linear combination of the one-body operators 
$
\frac{1}{2}(
\e^{ i\alpha (\hat{n}_{i\uparrow}+\hat{n}_{i\downarrow}-1)}+
\e^{-i\alpha (\hat{n}_{i\uparrow}+\hat{n}_{i\downarrow}-1)}
)
$ 
shown in Fig.~\ref{fig:LCU} 
can also be expressed with 
controlled-controlled-unitary gates, as shown in Fig.~\ref{fig:8x8}.
In Fig.~\ref{fig:8x8},
$\bs{0}$ is the $2\times 2$ null matrix,
$\bs{I}$ is the $2\times 2$ identity matrix,
and the Hadamard, $Z$-rotation, and $X$-rotation matrices
are given by
\begin{alignat}{1}
  &\bs{H}=
  \frac{1}{\sqrt{2}}
  \begin{bmatrix}
    1 & 1 \\
    1 & -1
  \end{bmatrix}, \\
  &\bs{R}_Z(\alpha)=
  \begin{bmatrix}
    \e^{-\imag \alpha/2} & 0 \\
    0 & \e^{\imag \alpha/2}
  \end{bmatrix},
\end{alignat}
and
\begin{alignat}{1}
  \bs{R}_X(\alpha)=
  \begin{bmatrix}
    \cos{\frac{\alpha}{2}} & -\imag \sin{\frac{\alpha}{2}}  \\
    -\imag \sin{\frac{\alpha}{2}} & \cos{\frac{\alpha}{2}}
  \end{bmatrix},
\end{alignat}
respectively. 
The resulting quantum circuit
using the controlled-controlled-$R_X$ gates
has an intuitive interpretation similar to Ref.~\cite{Murta2021}; 
Let us expand $|\psi_0\rangle$ by the computational-basis states
$\{|b\rangle\}_{b=0}^{2^{2N_{\rm site}-1}}$
as $|\psi_0\rangle=\sum_{b}\langle b|\psi_0\rangle |b\rangle$,
where $b$ denotes a bit string of length $2N_{\rm site}$. 
The operator 
$
\frac{1}{2}(
\e^{ i\alpha (\hat{n}_{i\uparrow}+\hat{n}_{i\downarrow}-1)}+
\e^{-i\alpha (\hat{n}_{i\uparrow}+\hat{n}_{i\downarrow}-1)}
)
=\cos{(2\alpha\hat{\eta}_i^z)}
$
applied to $|\psi_0\rangle$ then 
multiplies a factor $\cos{\alpha}=\e^{-g/2}$
[corresponding to the diagonal element of $\bs{R}_X(\pm 2\alpha)$]
to basis states $\{|b\rangle\}$
if they are in  
either doubly occupied configuration $|1\rangle_{i_\uparrow}|1\rangle_{i_\downarrow}$ 
or empty configuration $|0\rangle_{i_\uparrow}|0\rangle_{i_\downarrow}$  
(corresponding to the controlled-controlled parts),
while it
multiplies $1$ to the other basis states.

The second approach introduced here can be implemented in a much simpler quantum circuit 
composed essentially of single-qubit rotation ($R_Z$) gates without ancillary qubits. 
However, this approach cannot prepare the Gutzwiller wave function $|\psi_g\rangle$ itself
on a quantum computer. Rather, the expectation values of
observables are calculated stochastically by the importance sampling. 
Moreover, in general, the second approach suffers from
the phase problem, as does the auxiliary-field quantum Monte Carlo method in classical computation. 
However, 
when the phase problem is absent, 
the second approach can avoid the exponentially hard scaling in the first approach 
where the success probability $p_{00\cdots0}$ for preparing the Gutzwiller wave function decreases exponentially 
in the system size $N_{\rm site}$.
It should be noted that, 
in the absence of the phase problem,
the computational complexity of 
the auxiliary-field quantum Monte Carlo method
in classical computation is already
polynomial in $N_{\rm site}$, thus accessible currently to several hundreds to thousands of sites 
for Fermi-Hubbard-type models~\cite{Sorella2012,Otsuka2016,Otsuka2018,Seki2019Dirac,Otsuka2020,ulybyshev2021bridging},
and hence the quantum advantage of the second approach 
is not obvious.
However, unlike the auxiliary-field quantum Monte Carlo method, 
the present approach can be extended to a trial state $|\psi_0 \rangle$
that is not an uncorrelated Slater determinant state 
but a correlated multideterminant state,
the latter being prepared, 
for example, with a variational-quantum-eigensolver scheme.
Extension of the present approach to this direction will be
worth studying in the future. 
If the trial state $|\psi_0\rangle$ is spin separable, 
the corresponding quantum circuit further simplifies as it requires only  
$N_{\rm site}+1$ qubits, as compared to 
$2N_{\rm site}+1$ qubits for the spin-entangled trial state $|\psi_0\rangle$,
where ``$+1$'' qubit is the ancillary qubit for the Hadamard test.

The present scheme 
based on the discrete Hubbard-Stratonovich transformation
is somewhat similar to the recently proposed method
of decomposing a two-qubit unitary gate as a sum of single-qubit gates~\cite{Mitarai2021}
in the sense that the
Gutzwiller factor, corresponding to the two-body interaction, is 
decomposed into a product of the linear combination of unitary operators, corresponding
to one-body terms. 
A major difference from Ref.~\cite{Mitarai2021} is 
that the Gutzwiller factor is nonunitary and hence
no counterparts of the two-qubit unitary gate exist. 

The scheme proposed here has several extensions. 
A straightforward extension is to increase the number of variational parameters by
allowing  $g$ to be site dependent, i.e., $g\mapsto g_i$, under which
the rotation angle $\alpha$ in Eq.~(\ref{eq:alpha}) becomes site dependent
as $\alpha \mapsto \alpha_i=\arccos{(\e^{-g_i/2})}$. 
We also note that the site-dependent chemical potential or ``fugacity'' factors
in the Gutzwiller factor~\cite{Gebhard1990} can also be included
if a generalization of the discrete Hubbard-Stratonovich transformation~\cite{seki2019}
is employed. 
It is also possible to extend the Gutzwiller factor
to the Jastrow operator, which takes into account
long-range density-density correlations~\cite{Tang2018}, and 
to imaginary-time-evolution operators for electron-phonon-coupled systems~\cite{Karakuzu2018,Costa2020,Costa2021}, 
by using different kinds of Hubbard-Stratonovich transformations. 
A general framework for
obtaining discrete Hubbard-Stratonovich transformations
of the exponentiated density-density interactions is provided in Appendix~\ref{app:A}.
The research along this line is now in progress. 

\acknowledgements
A part of the numerical simulations has been performed
using the HOKUSAI supercomputer at RIKEN
(Project ID: Q21532, ID Q21525, and ID Q22525) and also supercomputer Fugaku installed in RIKEN R-CCS.
This work is supported by
Grant-in-Aid for Research Activity start-up (No.~JP19K23433),
Grant-in-Aid for Scientific Research (C) (No.~JP18K03475, No.~JP21K03395, and No.~JP22K03520),
Grant-in-Aid for Scientific Research (B) (No.~JP18H01183), and 
Grant-in-Aid for Scientific Research (A) (No.~JP21H04446) from MEXT, Japan. 
This work is also supported in part by the COE research grant in computational science from 
Hyogo Prefecture and Kobe City through Foundation for Computational Science.

\appendix

\section{Discrete Hubbard-Stratonovich transformations}~\label{app:A}

In this Appendix,  
we provide a general framework for obtaining 
Hubbard-Stratonovich transformations that transform 
$\e^{-J \hat{Z}_i \hat{Z}_j}$ as a linear combination of
unitary operators, both for $J<0$ and $J>0$. 
Since $(\hat{n}_{i\sigma}-\frac{1}{2})(\hat{n}_{j\sigma'}-\frac{1}{2}) 
\overset{\rm JWT}{=} \frac{1}{4}\hat{Z}_{i_\sigma}\hat{Z}_{j_{\sigma'}}$, 
this implies that 
any exponentiated density-density interactions can be decomposed into a linear combination of unitary operators.

\subsection{$\exp{(-J\hat{Z}_i\hat{Z}_j)}$ with $J<0$}

We consider the case of $J=-|J|<0$. 
We begin with the matrix representation 
\begin{equation}
  \e^{-J\hat{Z}_i \hat{Z}_j}
  \overset{\cdot}{=}
  \begin{bmatrix}
    \e^{-J} & 0 & 0 & 0 \\
    0 & \e^{J} & 0 & 0   \\
    0 & 0 & \e^{J} & 0   \\
    0 & 0 & 0 & \e^{-J} 
  \end{bmatrix},
  \label{eq:mat_zz}
\end{equation}
where $\e^{-J} > 1$ and $\e^{J} < 1$ because $J<0$. 
Next, according to Ref.~\cite{seki2021spatial}, 
typical two-qubit two-level unitaries have the matrix representations 
\begin{alignat}{1}
  & \exp\left[-\imag \frac{\alpha}{2} \left(\hat{X}_i \hat{X}_j + \hat{Y}_i \hat{Y}_j\right) \right] \overset{\cdot}{=}
  \begin{bmatrix}
    1 & 0 & 0 & 0 \\
    0 & \cos{\alpha} & -\imag\sin{\alpha} & 0 \\
    0 & -\imag\sin{\alpha} & \cos{\alpha} & 0 \\
    0 & 0 & 0 & 1
  \end{bmatrix},
  \label{expxmat}\\
  & \exp\left[-\imag \frac{\alpha}{2} \left(\hat{X}_i \hat{Y}_j - \hat{Y}_i \hat{X}_j\right) \right] \overset{\cdot}{=}
  \begin{bmatrix}
    1 & 0 & 0 & 0 \\
    0 & \cos{\alpha} & \sin{\alpha} & 0 \\
    0 & -\sin{\alpha} & \cos{\alpha} & 0 \\
    0 & 0 & 0 & 1
  \end{bmatrix}, \label{expymat} \\
  &
  \exp\left[-\imag \frac{\alpha}{2} \left(\hat{Z}_j - \hat{Z}_i \right) \right] \overset{\cdot}{=}
  \begin{bmatrix}
    1 & 0 & 0 & 0 \\
    0 & \e^{\imag \alpha} & 0 & 0 \\
    0 & 0 & \e^{-\imag \alpha} & 0 \\
    0 & 0 & 0 & 1
  \end{bmatrix}, \label{expzmat}
\end{alignat}
where $\alpha$ is real. 
Since $\sin{\alpha}$ ($\cos{\alpha}$) is an odd (even) function of $\alpha$, 
linear combinations of these two-qubit two-level unitaries with opposite rotation angles
result in a diagonal matrix, i.e.,  
\begin{alignat}{1}
  & \sum_{s=\pm1}\exp\left[-\imag \frac{s\alpha}{2} \left(\hat{X}_i \hat{X}_j + \hat{Y}_i \hat{Y}_j\right) \right] \notag  \\
  =& \sum_{s=\pm1}\exp\left[-\imag \frac{s\alpha}{2} \left(\hat{X}_i \hat{Y}_j - \hat{Y}_i \hat{X}_j\right) \right] \notag  \\
  =& \sum_{s=\pm1}\exp\left[-\imag \frac{s\alpha}{2} \left(\hat{Z}_j - \hat{Z}_i \right) \right] \notag  \\
  \overset{\cdot}{=}&
  \begin{bmatrix}
    2 & 0 & 0 & 0 \\
    0 & 2\cos{\alpha} & 0 & 0   \\
    0 & 0 & 2\cos{\alpha} & 0   \\
    0 & 0 & 0 & 2
  \end{bmatrix}.
  \label{eq:mat_hst1}
\end{alignat}
Therefore, 
comparing Eqs.~(\ref{eq:mat_zz}) and (\ref{eq:mat_hst1}), 
we find the discrete Hubbard-Stratonovich transformations
\begin{alignat}{1}
  \e^{-J\hat{Z}_i \hat{Z}_j}
  =&\gamma \sum_{s=\pm1}\exp\left[-\imag \frac{s\alpha}{2} \left(\hat{X}_i \hat{X}_j + \hat{Y}_i \hat{Y}_j\right) \right] \label{hst_Givens_X}\\
  =&\gamma \sum_{s=\pm1}\exp\left[-\imag \frac{s\alpha}{2} \left(\hat{X}_i \hat{Y}_j - \hat{Y}_i \hat{X}_j\right) \right] \label{hst_Givens_Y}\\
  =&\gamma \sum_{s=\pm1}\exp\left[-\imag \frac{s\alpha}{2} \left(\hat{Z}_j - \hat{Z}_i \right) \right] \label{hst_Givens_Z}
\end{alignat}
with $\gamma=\e^{-J}/2$ and $\alpha=\arccos{(\e^{2J})}$.

\subsection{$\exp{(-J\hat{Z}_i\hat{Z}_j)}$ with $J>0$}

Next, we consider the case of $J>0$,  
for which $\e^{-J} < 1$ and $\e^{J} > 1$ in Eq.~(\ref{eq:mat_zz}). 
According to Ref.~\cite{seki2021spatial}, 
typical two-qubit two-level unitaries have the matrix representations 
\begin{alignat}{1}
  & \exp\left[-\imag \frac{\alpha}{2} \left(\hat{X}_i \hat{X}_j - \hat{Y}_i \hat{Y}_j\right) \right] \overset{\cdot}{=}
  \begin{bmatrix}
    \cos{\alpha} & 0 & 0 & -\imag\sin{\alpha} \\
    0 & 1 & 0 & 0 \\
    0 & 0 & 1 & 0 \\
    -\imag\sin{\alpha} & 0 & 0 & \cos{\alpha} \\
  \end{bmatrix}, \label{expxmat2}\\
  & \exp\left[-\imag \frac{\alpha}{2} \left(\hat{X}_i \hat{Y}_j + \hat{Y}_i \hat{X}_j\right) \right] \overset{\cdot}{=}
  \begin{bmatrix}
    \cos{\alpha} & 0 & 0 & -\sin{\alpha}  \\
    0 & 1 & 0 & 0 \\
    0 & 0 & 1 & 0 \\
    \sin{\alpha} & 0 & 0 & \cos{\alpha}  \\
  \end{bmatrix}, \label{expymat2}\\
  &\exp\left[-\imag \frac{\alpha}{2} \left(\hat{Z}_j + \hat{Z}_i \right) \right] \overset{\cdot}{=}
  \begin{bmatrix}
    \e^{-\imag \alpha} & 0 & 0 & 0 \\
    0 & 1 & 0 & 0 \\
    0 & 0 & 1 & 0 \\
    0 & 0 & 0 & \e^{\imag \alpha}
  \end{bmatrix}, \label{expzmat2}
\end{alignat}
where $\alpha$ is real.  
Since $\sin{\alpha}$ ($\cos{\alpha}$) is an odd (even) function of $\alpha$, 
linear combinations of these two-qubit two-level unitaries with opposite rotation angles
result in a diagonal matrix, i.e., 
\begin{alignat}{1}
   & \sum_{s=\pm1}\exp\left[-\imag \frac{s\alpha}{2} \left(\hat{X}_i \hat{X}_j - \hat{Y}_i \hat{Y}_j\right) \right] \notag  \\
  =& \sum_{s=\pm1}\exp\left[-\imag \frac{s\alpha}{2} \left(\hat{X}_i \hat{Y}_j + \hat{Y}_i \hat{X}_j\right) \right] \notag  \\
  =& \sum_{s=\pm1}\exp\left[-\imag \frac{s\alpha}{2} \left(\hat{Z}_j + \hat{Z}_i \right) \right] \notag  \\
  \overset{\cdot}{=}&
  \begin{bmatrix}
    2\cos{\alpha} & 0 & 0 & 0 \\
    0 & 2 & 0 & 0   \\
    0 & 0 & 2 & 0   \\
    0 & 0 & 0 & 2\cos{\alpha}
  \end{bmatrix}.
  \label{eq:mat_hst2}
\end{alignat}
Therefore, 
comparing Eqs.~(\ref{eq:mat_zz}) and (\ref{eq:mat_hst2}), 
we find the discrete Hubbard-Stratonovich transformations
\begin{alignat}{1}
  \e^{-J\hat{Z}_i \hat{Z}_j}
  =&\gamma \sum_{s=\pm1}\exp\left[-\imag \frac{s\alpha}{2} \left(\hat{X}_i \hat{X}_j - \hat{Y}_i \hat{Y}_j\right) \right] \label{hst_Bogoliubov_X} \\
  =&\gamma \sum_{s=\pm1}\exp\left[-\imag \frac{s\alpha}{2} \left(\hat{X}_i \hat{Y}_j + \hat{Y}_i \hat{X}_j\right) \right] \label{hst_Bogoliubov_Y} \\
  =&\gamma \sum_{s=\pm1}\exp\left[-\imag \frac{s\alpha}{2} \left(\hat{Z}_j + \hat{Z}_i \right) \right] \label{hst_Bogoliubov_Z}
\end{alignat}
with $\gamma=\e^{J}/2$ and $\alpha=\arccos{(\e^{-2J})}$. 

Equation~(\ref{hst_Bogoliubov_Z})
corresponds to the discrete Hubbard-Stratonovich transformation
in Ref.~\cite{Hirsch1983} and is adopted here in this study,  
while Eqs.~(\ref{hst_Bogoliubov_X}) and (\ref{hst_Bogoliubov_Y})
are different discrete Hubbard-Stratonovich transformations, which might be 
useful for other purposes. 
Finally, we note that 
Eqs.~(\ref{hst_Bogoliubov_X}) and (\ref{hst_Bogoliubov_Y}) 
are similar to the discrete Hubbard-Stratonovich transformation for fermions 
in anomalous channels discussed in Ref.~\cite{Batrouni1990}.

\section{Absence of the phase problem} \label{sec:pp}
In this Appendix, we prove that the phase problem is absent for our particular case, i.e., 
$|\psi_0\rangle$ being the ground state of $\hat{K}$ on a bipartite lattice composed of sublattices $A$ and $B$ at half filling.
For this purpose, we introduce a unitary operator 
$\hat{\cal U}_{\rm pPH}$ for the partial particle-hole (pPH) transformation such that 
\begin{alignat}{1}
  &\hat{\cal U}_{\rm pPH} \hat{c}_{i \uparrow} \hat{\cal U}_{\rm pPH}^{-1} = \hat{c}_{i \uparrow}, \label{eq:pPH1}\\
  &\hat{\cal U}_{\rm pPH} \hat{c}_{i \downarrow} \hat{\cal U}_{\rm pPH}^{-1} = (-1)^i\hat{c}_{i \downarrow}^\dag \label{eq:pPH2}, 
\end{alignat}
where $(-1)^i$ takes
the different sign when site $i$ belongs to the different sublattice on the bipartite lattice.
We assume that the number of sites is even.
Since 
$\hat{\cal U}_{\rm pPH}\hat{K}\hat{\cal U}_{\rm pPH}^{-1} =\hat{K}$ and 
$\hat{\cal U}_{\rm pPH}\hat{D}\hat{\cal U}_{\rm pPH}^{-1} = -\hat{D}$, 
$\hat{K}$ is invariant but $\hat{\cal H}$ is not invariant under the pPH transformation.
Rather, the pPH transformation transforms
the repulsive Fermi-Hubbard model to the attractive Fermi-Hubbard model~\cite{Shiba1972}.
An explicit form of $\hat{\cal U}_{\rm pPH}$ can be written as
$\hat{\cal U}_{\rm pPH}=\prod_{i=1}^{N_{\rm site}}
(\hat{c}_{i\downarrow}+(-1)^i\hat{c}_{i\downarrow}^\dag)$
(see for example Refs.~\cite{Hatsugai2006,Tasaki}).

We also introduce an antiunitary operator
$\hat{\cal A}_{\rm TR}$ for the time-reversal (TR) operation such that 
\begin{alignat}{1}
  &\hat{\cal A}_{\rm TR} \hat{c}_{i \uparrow} \hat{\cal A}_{\rm TR}^{-1} = \hat{c}_{i \downarrow}, \label{eq:TR1}\\
  &\hat{\cal A}_{\rm TR} \hat{c}_{i \downarrow} \hat{\cal A}_{\rm TR}^{-1} = -\hat{c}_{i \uparrow}, \label{eq:TR2} 
\end{alignat}
and similarly for $\hat{c}_{i\sigma}^\dag$.
The TR operator $\hat{\cal A}_{\rm TR}$ can be written as 
$\hat{\cal A}_{\rm TR}=\hat{\cal U}_{\rm TR}\hat{C}$,
where $\hat{\cal U}_{\rm TR}$ is a unitary operator and $\hat{C}$ is the
complex-conjugation operator, and hence 
\begin{equation}
  \hat{\cal A}_{\rm TR} z \hat{\cal A}_{\rm TR}^{-1}=z^* \label{eq:cc}
\end{equation}
for any complex number $z$.
We note that the unitary part $\hat{\cal U}_{\rm TR}$ of the
TR operator can be explicitly written as
$\hat{\cal U}_{\rm TR}=\prod_{i=1}^{N_{\rm site}}
\hat{\cal F}_{i\uparrow,i\downarrow} \e^{\imag \pi \hat{n}_{i\downarrow}}$,
where $\hat{\cal F}_{i\sigma,j\sigma'}=
1+(\hat{c}_{i\sigma}^\dag \hat{c}_{j\sigma'} + {\rm H.c.})
-\hat{c}_{i\sigma}^\dag \hat{c}_{i\sigma}
-\hat{c}_{j\sigma'}^\dag \hat{c}_{j\sigma'}$
is the fermionic SWAP operator~\cite{Bravyi2002,Essler2005,Verstraete2009,Barthel2009,Wecker2015,Kivlichan2018,seki2021spatial} 
and $\e^{\imag \phi \hat{n}_{i\downarrow}}=1+(\e^{\imag \phi}-1)\hat{n}_{i\downarrow}
\overset{\phi=\pi}{=}1-2\hat{n}_{i\downarrow}$
accounts for the gauge transformation~\cite{Tasaki}
for the spin-down fermions in Eq.~(\ref{eq:TR2}).
From the properties of the fermionic SWAP operator~\cite{seki2021spatial} 
it can be confirmed that 
$\hat{\cal A}_{\rm TR}^2=\prod_{i=1}^{N_{\rm sites}}(-1)^{\hat{n}_{i\uparrow}+\hat{n}_{i\downarrow}}=(-1)^{\hat{N}}$. 
Since 
$\hat{\cal A}_{\rm TR}\hat{K}\hat{\cal A}_{\rm TR}^{-1} =\hat{K}$ and 
$\hat{\cal A}_{\rm TR}\hat{D}\hat{\cal A}_{\rm TR}^{-1} =\hat{D}$,
$\hat{K}$, $\hat{D}$, and hence $\hat{\cal H}$ are invariant under the TR operation.

For the later convenience, we also introduce another 
antiunitary operator $\hat{\Theta}$ as
$\hat{\Theta}=\hat{\cal U}_{\rm pPH}\hat{\cal A}_{\rm TR}$. 
It follows from $\hat{\Theta}\hat{K}\hat{\Theta}^{-1}=\hat{K}$ that,  
if $|\psi_0\rangle$ is an eigenstate of $\hat{K}$,
then $|\tilde{\psi}_0\rangle \equiv \hat{\Theta}|\psi_0\rangle$ is
also an eigenstate of $\hat{K}$ with the same eigenvalue.
It should be reminded that, 
if $\hat{\cal A}$ is an antiunitary operator, 
$|\psi\rangle$ and $|\phi\rangle$ are some states, and
$|\tilde{\psi}\rangle \equiv \hat{\cal A}|\psi\rangle$ and
$|\tilde{\phi}\rangle \equiv \hat{\cal A}|\phi\rangle$
are the antiunitary-operated states associated with $|\psi\rangle$ and $|\phi\rangle$, respectively, 
then a matrix element $\langle \psi|\hat{X}|\phi\rangle$ of
a linear operator $\hat{X}$
can be written in terms of 
$|\tilde{\psi}\rangle$ and $|\tilde{\phi}\rangle$ as~\cite{Sakurai,batanouny_wooten_2008}
\begin{equation}
  \langle \psi|\hat{X}|\phi\rangle= \langle \tilde{\phi}|\hat{\cal A} \hat{X}^\dag  \hat{\cal A}^{-1} |\tilde{\psi}\rangle.
  \label{eq:me_au}
\end{equation}

Now we consider how the operator
$\hat{u}_{\bs{s},\sigma}\equiv \prod_\tau \hat{u}_{\bs{s}_\tau, \sigma}$ 
for a given set of auxiliary fields $\bs{s}=\{s_{i,\tau}\}$
is transformed by
$\hat{\cal U}_{\rm pPH}$, $\hat{\cal A}_{\rm TR}$, and $\hat{\Theta}$.  
From Eqs.~(\ref{eq:pPH1}) and (\ref{eq:pPH2}), it follows that 
\begin{equation}
  \hat{\cal U}_{\rm pPH}
  (\hat{u}_{\bs{s},\uparrow} \otimes \hat{u}_{\bs{s},\downarrow})
  \hat{\cal U}_{\rm pPH}^{-1} =
  \hat{u}_{\bs{s},\uparrow} \otimes \hat{u}_{\bs{s},\downarrow}^*.
  \label{eq:pPHu}
\end{equation}
From Eqs.~(\ref{eq:TR1}), (\ref{eq:TR2}) and (\ref{eq:cc}), it follows that 
\begin{equation}
  \hat{\cal A}_{\rm TR}
  (\hat{u}_{\bs{s},\uparrow} \otimes \hat{u}_{\bs{s},\downarrow})
  \hat{\cal A}_{\rm TR}^{-1} =
  \hat{u}_{\bs{s},\uparrow}^* \otimes \hat{u}_{\bs{s},\downarrow}^*.
  \label{eq:TRu}
\end{equation}
From Eqs.~(\ref{eq:pPHu}) and (\ref{eq:TRu}), it follows that
\begin{equation}
  \hat{\Theta}
  (\hat{u}_{\bs{s},\uparrow} \otimes \hat{u}_{\bs{s},\downarrow})
  \hat{\Theta}^{-1}=
  \hat{u}_{\bs{s},\uparrow}^* \otimes \hat{u}_{\bs{s},\downarrow}.
  \label{eq:theta}
\end{equation}
We also note that $\hat{u}_{\bs{s},\sigma}^*=\hat{u}_{\bs{s},\sigma}^\dag$ because
$[\hat{u}_{\bs{s}_\tau,\sigma},\hat{u}_{\bs{s}_{\tau'},\sigma}]=0$.

Next we consider the matrix element
\begin{equation}
  W \equiv \langle \psi_0|
  \hat{u}_{\bs{s},\uparrow} \otimes \hat{u}_{\bs{s},\downarrow}
  |\psi_0\rangle,
\end{equation}
which corresponds to the numerator of $P_{\bs{s}}$ in Eq.~(\ref{eq:ps}). 
Here, we assume that 
$|\psi_0\rangle$ is the unique ground state
of $\hat{K}$, implying that
$|\tilde{\psi}_0\rangle = \hat{\Theta}|\psi_0\rangle$ differs from $|\psi_0\rangle$ only
by a phase factor, $|\tilde{\psi}_0 \rangle= \e^{\imag \theta}|\psi_0\rangle$. 
Note also that $|\psi_0\rangle$ is a spin-separable state.
Then, by noticing that $\hat{\Theta}$ is an antiunitary operator,
$W$ can be written as
\begin{alignat}{1}
  W
  &= \langle \tilde{\psi}_0|
  \hat{\Theta} (\hat{u}_{\bs{s},\uparrow} \otimes \hat{u}_{\bs{s},\downarrow})^\dag \hat{\Theta}^{-1} 
  |\tilde{\psi}_0\rangle \notag \\
  &= \langle \tilde{\psi}_0|
  ((\hat{u}_{\bs{s},\uparrow}^\dag)^* \otimes \hat{u}_{\bs{s},\downarrow}^\dag)
  |\tilde{\psi}_0\rangle \notag \\
  &=
  \langle \psi_{0,\uparrow} | (\hat{u}_{\bs{s},\uparrow}^*)^\dag|\psi_{0,\uparrow}\rangle
  \langle \psi_{0,\downarrow} | \hat{u}_{\bs{s},\downarrow}^\dag|\psi_{0,\downarrow}\rangle \notag \\
  &\equiv W_\uparrow W_\downarrow.
  \label{eq:W}
\end{alignat}
Here, we have used Eq.~(\ref{eq:me_au}) in the first equality and Eq.~(\ref{eq:theta}) in the second equality. 
Since 
$\hat{\cal A}_{\rm TR}\hat{K}_\uparrow \hat{\cal A}_{\rm TR}^{-1}=\hat{K}_{\downarrow}$ and 
$\hat{\cal A}_{\rm TR}\hat{K}_\downarrow \hat{\cal A}_{\rm TR}^{-1}=\hat{K}_{\uparrow}$,
we assume that $|\psi_{0,\downarrow}\rangle$ is the time-reversed state of $|\psi_{0,\uparrow}\rangle$,
i.e., $|\psi_{0,\downarrow}\rangle=\hat{\cal A}_{\rm TR}|\psi_{0,\uparrow}\rangle$
up to a phase factor. 
Then, by noticing that $\hat{\cal A}_{\rm TR}$ is antiunitary,
$W_{\uparrow}$ in Eq.~(\ref{eq:W}) can be written as
\begin{alignat}{1}
  W_\uparrow
  &=
  \langle \psi_{0,\downarrow}|\hat{\cal A}_{\rm TR} \hat{u}_{\bs{s},\uparrow}^* \hat{\cal A}_{\rm TR}^{-1}|\psi_{0,\downarrow}\rangle \notag \\
  &=
  \langle \psi_{0,\downarrow}|\hat{u}_{\bs{s},\downarrow} |\psi_{0,\downarrow}\rangle \notag \\
  &= W_{\downarrow}^*. 
\end{alignat}
Therefore, $W=|W_\downarrow|^2>0$, proving that the phase problem is absent in this case. 
This is essentially the same argument used to prove the absence of the negative sign problem 
for the Fermi-Hubbard model in the auxiliary-field quantum Monte-Carlo method~\cite{Hirsch1985Hub,Congjun2005,Zheng2011}.

\bibliography{biball}

\begin{thebibliography}{118}
\bibitem [{not()}]{note_decomp}%
  \BibitemOpen
  \href@noop {} {}\bibinfo {note} {Note however that $\e^{-2g\hat{D}}$ in the
  denominator in Eq.~(\ref{eq:aveO}) can also be decomposed directly without
  introducing the two kinds of auxiliary fields $\{s_{i,\tau} \}$ with
  $\tau=1,2$.}\BibitemShut {Stop}%
\end{thebibliography}
\end{document}